\documentclass[prb,twocolumn,superscriptaddress]{revtex4-1}

\usepackage{amsmath,color,amsfonts,wasysym,graphicx,MnSymbol}

\newcommand{\ket}[1]{\vert#1\rangle}
\newcommand{\bra}[1]{\langle#1\vert}
\newcommand{\lrket}[1]{\left\vert#1\right\rangle}
\newcommand{\ketbra}[1]{\ket{#1}\!\bra{#1}}
\newcommand{\mc}[1]{\mathcal{#1}}
\newcommand{\mf}[1]{\mathfrak{#1}}
\newcommand{\mb}[1]{\mathbb{#1}}
\newcommand{\ds}{\davidsstar}
\newcommand{\ZZ}{\bar{Z}}
\newcommand{\rg}{\mathrm{rg}}

\newcommand{\Hc}{{\cal H}}
\newcommand{\Z}{\mathbb{Z}}
\renewcommand{\>}{\rangle}

\begin{document}

\title{Resonating valence bond states in the PEPS formalism}

\author{Norbert Schuch}
\affiliation{Institut f\"ur Quanteninformation, 
    RWTH Aachen, 52056 Aachen, Germany}
\affiliation{Institute for Quantum Information, California Institute of
Technology, MC 305-16, Pasadena CA 91125, U.S.A.}
\author{Didier Poilblanc}
\affiliation{Laboratoire de Physique Th\'eorique, C.N.R.S.\ 
    and Universit\'e de Toulouse, 31062 Toulouse, France}
\author{J.~Ignacio \surname{Cirac}}
\affiliation{Max-Planck-Institut f\"ur Quantenoptik,
Hans-Kopfermann-Str.\ 1, D-85748 Garching, Germany}
\author{David P\'erez-Garc\'ia}
\affiliation{Dpto.\ Analisis Matematico and IMI, 
    Universidad Complutense de Madrid, 
    E-28040 Madrid, Spain}

\begin{abstract}
We study resonating valence bond (RVB) states in the Projected Entangled
Pair States (PEPS) formalism.  Based on symmetries in the PEPS
description, we establish relations between the toric code state, the
orthogonal dimer state, and the $\mathrm{SU}(2)$ singlet RVB state on the
kagome lattice: We prove the equivalence of toric code and dimer state,
and devise an interpolation between the dimer state and the RVB
state. This interpolation corresponds to a continuous path in Hamiltonian
space, proving that the RVB state is the four-fold degenerate ground state
of a local Hamiltonian on the (finite) kagome lattice.  We investigate this
interpolation using numerical PEPS methods, studying the decay of correlation
functions, the change of overlap, and the entanglement spectrum, none of
which exhibits signs of a phase transition.
\end{abstract} 

\maketitle


\section{Introduction}

Resonating valence bond (RVB) states have been introduced by
Anderson~\cite{anderson:rvb} as a wavefunction in the context of
high-temperature superconductivity and have since then received
significant attention.  RVB wavefunctions are obtained as the
superposition of all nearest-neighbor (or otherwise constrained) singlet
coverings on a given lattice, and they have been studied as an ansatz
capturing the behavior of frustrated quantum spin systems, in particular
Heisenberg antiferromagnets on frustrated
lattices.~\cite{misguich:AFM-review}    They are believed to describe
so-called \emph{topological spin liquids}, i.e., exotic phases with
degenerate ground states which however do not break rotational or
translational symmetry, despite the presence of strong antiferromagnetic
interactions.  Particular interest, both theoretically and experimentally,
has been devoted to frustrated magnets on the kagome lattice, as those
systems are realized in actual materials and form candidates for the first
experimental observation of a topological spin
liquid.~\cite{mendels:kagome,meng:spinliquid-2d-diracfermions,%
mezzacapo:honeycomb-j1j2,yan:heisenberg-kagome,%
depenbrock:kagome-heisenberg-dmrg}

Unfortunately, models with potential RVB ground states are typically
difficult to study. One the one hand, this is due to presence of
frustration. Yet, the bigger issue seems
to be that different singlet configurations are not orthogonal, which up
to now has e.g.\ hindered a full understanding of how RVB states appear
as ground states of local Hamiltonians.~\cite{seidel:kagome}  In order to
better understand the RVB phase, simplified so-called \emph{dimer models}
have been introduced,~\cite{rokhsar:dimer-models} where the system is
re-defined on the Hilbert space spanned by all dimer coverings of the
lattice, making different dimer configurations orthogonal by definition.
Dimer models have subsequently been studied for various lattices, and it
has in particular been found that for certain geometries, such as for the
triangular~\cite{moessner:dimer-triangular} or the kagome
lattice,~\cite{misguich:dimer-kagome} such states appear as ground states
of local Hamiltonians with topological ground state degeneracy (four-fold
on the torus), forming a $\mathbb Z_2$ topological spin liquid.
Unfortunately, these findings cannot be mapped back to the true RVB state
due to the different underlying Hilbert spaces which are not related by a
simple mapping, and thus, the way in which the corresponding RVB states
can describe ground states of local Hamiltonians is only partially
understood (although very remarkable progress in this direction has been
made during the last years~\cite{cano:rvb-hamiltonians,seidel:kagome}).

More recently, Projected Entangled Pair States (PEPS) have been introduced
as a tool to study quantum many-body
wavefunctions.~\cite{verstraete:mbc-peps,verstraete:2D-dmrg} PEPS give a
description of quantum many-body states based on their entanglement
structure in terms of local tensors, and form a framework which allows to
both analytically understand these wavefuctions, and to study their
properties numerically.  In particular, there is a clear way to understand
how PEPS arise as ground states of local Hamiltonians, especially for
systems with unique ground states~\cite{perez-garcia:parent-ham-2d} and
with topological order,~\cite{schuch:peps-sym} based on the symmetry
properties of the underlying tensor representation. Numerically, they form
a tool for computing quantities such as correlation functions or overlaps
efficiently with very high accuracy, using transfer operators of matrix
product form.~\cite{verstraete:2D-dmrg}

In this paper, we perform a systematic study of RVB and dimer states in
the PEPS formalism, focusing on the kagome lattice.  We introduce closely
related PEPS representations for the RVB
wavefunction~\cite{verstraete:comp-power-of-peps} and a version of the
dimer state in which different dimer configurations are locally
orthogonal.  These representations allow us to derive a reversible local
mapping between the RVB and the dimer state, as well as to prove local
unitary equivalence of the dimer state and Kitaev's toric
code.\cite{kitaev:toriccode} This yields a Hamiltonian for our dimer
model, and subsequently a Hamiltonian for the RVB state: This is, we can
for the first time prove that the RVB state on the kagome lattice is the
ground state of a local Hamiltonian with topological ground state
degeneracy (four-fold on the torus) for any finite lattice.  The PEPS
formalism further allows us to construct an interpolation between RVB and
dimer state where the  Hamiltonian changes smoothly along the path.  Using
techniques for numerical PEPS calculations,~\cite{verstraete:2D-dmrg} we
try to assess whether the RVB state is in the same phase as the dimer
state (which is equivalent to the toric code and thus a $\mathbb Z_2$
topological spin liquid) by looking for signatures of a phase transition along
the dimer--RVB interpolation.  We have considered the behavior of
correlation functions, the rate at which the ground state changes in terms
of the wave function overlap,\cite{zanardi:overlap-criterion} 
and the entanglement spectrum of the system, and have found that all of
these quantities behave smoothly and show no sign of a phase transition.

Let us describe the organization of the paper.  For the sake of
conciseness, alternative definitions, proofs, etc.\ have been moved to
appendices.  In Sec.~\ref{sec:definitions}, we introduce the RVB and dimer
states and their PEPS representations, as well as the formulation using
tensor networks.  Alternative PEPS representations are discussed in
Appendix~\ref{appendix:rvb-peps-representations}. In
Sec.~\ref{sec:relations}, we discuss the relations between the toric code,
the dimer state, and the RVB state, using the symmetry of the underlying
tensors.  Based on these findings, we show how to construct a smooth
interpolation between the dimer state and the RVB state in
Sec.~\ref{sec:interpolation}, and subsequently study its properties
numerically.  In Sec.~\ref{sec:hamiltonians}, we use the relation between
the toric code, the dimer state, and the RVB state to construct parent
Hamiltonians for the dimer state and the RVB, as well as a smooth path of
Hamiltonians for the interpolation between them.
Appendix~\ref{appendix:oRVB-parent} provides a simpler parent Hamiltonian
for the dimer state, based on a more direct mapping to the toric code, and
Appendix~\ref{appendix:twostars} shows how to directly derive a (much more
compact) parent Hamiltonian for the RVB state by generalizing the
techniques developed in Ref.~\onlinecite{schuch:peps-sym} for PEPS with
symmetries to the symmetries found in the RVB.

\section{Definitions
\label{sec:definitions}}

\subsection{The RVB and orthogonal RVB state}

\begin{figure}
\includegraphics[width=\columnwidth]{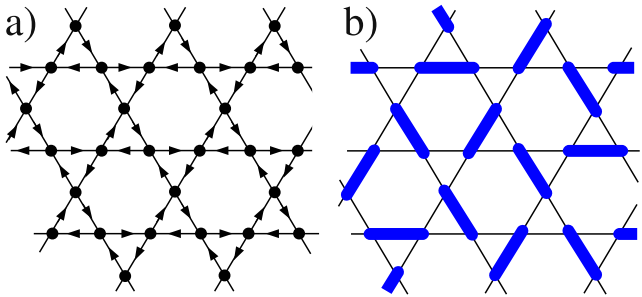}
\caption{
\label{fig:kagome-dimers}
\textbf{a)} Oriented kagome lattice. Spins are associated to vertices.
\textbf{b)} Dimer covering of the kagome lattice. The lattice is
completely covered with \emph{dimers} (marked blue), i.e., disjoint pairs
of adjacent vertices.  }
\end{figure}

We start by introducing dimer states and resonating valence bond (RVB)
states.  We focus on the kagome lattice, Fig.~\ref{fig:kagome-dimers}a,
both for its relevance and for clarity of the presentation; but our
techniques generalize to other lattices (see conclusions). 
 A \emph{dimer} is a pair of vertices connected by an edge. A
\emph{dimer covering} is a complete covering of the lattice with dimers,
Fig.~\ref{fig:kagome-dimers}b.  
 We can associate orthogonal quantum states $\ket D$ with each dimer
covering $D$.  Then, the \emph{dimer state} is given by the equal weight
superposition $\ket{\Psi_\mathrm{dimer}}=\sum\ket{D}$, where the sum runs
over all dimer coverings $D$.  Usually, it is favorable to ensure
orthogonality of different states $\ket{D}$ and $\ket{D'}$ locally; we
will introduce such a version of the dimer state which is particularly
suited for our purposes in Sec.~\ref{sec:dimer-peps-rep}.

Let us now turn towards the resonating valence bond (RVB) state. We first
associate to each vertex of the lattice a spin-$\tfrac12$ particle, or
qubit, with basis states $\ket{0}\equiv\lrket{\uparrow}$ and
$\ket{1}\equiv\lrket{\downarrow}$.  Then, for each dimer covering $D$
we define a state $\ket{\sigma(D)}$ which is a tensor product of singlets
$\ket{01}-\ket{10}$ (we omit normalization throughout) between the pairs
of spins in each dimer in the covering, where the singlets are oriented
according to the arrows in Fig.~\ref{fig:kagome-dimers}a. The
\emph{resonating valence bond} (RVB) state is then defined as the equal
weight superposition $\ket{\Psi_\mathrm{RVB}} = \sum_D \ket{\sigma(D)}$
over all dimer coverings.

\subsection{PEPS representation of the RVB state
\label{sec:RVB-peps-representation}
}

We will now give a description of the RVB state in terms of Projected
Entangled Pair States (PEPS).\cite{verstraete:mbc-peps} PEPS are states
which can be described by first placing ``virtual'' entangled states
between the sites of the system, and subsequently applying linear maps at
each site to obtain the physical system. A PEPS representation of RVB
states has first been given in
Ref.~\onlinecite{verstraete:comp-power-of-peps}; a detailed discussion how
it is related to our description can be found in
Appendix~\ref{appendix:rvb-peps-representations}.

\begin{figure}
\includegraphics[width=0.6\columnwidth]{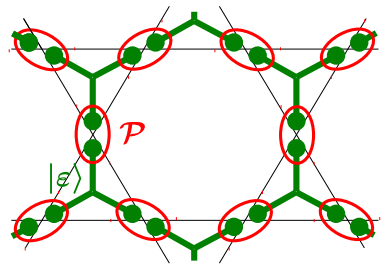}
\caption{\label{fig:rvb-eps-p2-peps}
PEPS construction of the RVB state:  Place $\ket\epsilon$ states
(green), Eq.~(\ref{eq:eps-bond}), and then apply the map $\mc P$ (red),
Eq.~(\ref{eq:rvb-proj}), as indicated.  To obtain the dimer state, replace
$\mc P$ by $\mc P_\perp$, Eq.~(\ref{eq:orvb-proj}).}
\end{figure}

To obtain a PEPS description of the RVB state, we first place $3$-qutrit
states
\begin{equation}
       \label{eq:eps-bond}
\ket{\varepsilon} = \sum_{i,j,k=0}^2 \varepsilon_{ijk}\ket{ijk} + \ket{222}\ ,
\end{equation}
inside each triangle of the kagome lattice, as depicted in
Fig.~\ref{fig:rvb-eps-p2-peps}. Here, $\varepsilon_{ijk}$ is the
completely antisymmetric tensor with $\varepsilon_{012}=1$, and $i$, $j$,
and $k$ are oriented clockwise (i.e., consistent with the arrows in
Fig.~\ref{fig:kagome-dimers}a).  Second, we apply the map
\begin{equation}
        \label{eq:rvb-proj}
\mc P = \ket{0}(\bra{02}+\bra{20})+\ket{1}(\bra{12}+\bra{21})
\end{equation}
at each vertex, which maps the two qutrits from the adjacent
$\ket{\varepsilon}$ states to one qubit.  It is straightforward to check
that this construction exactly gives the resonating valence bond state
defined above (see Appendix~\ref{appendix:rvb-peps-representations} for a
detailed discussion).

\subsection{PEPS representation of dimer state
\label{sec:dimer-peps-rep}}

In a similar way as for the RVB state, we can also obtain a PEPS
representation of the dimer state. To this end, we enlarge our local
Hilbert space and replace the map $\mc P$ in
Fig.~\ref{fig:rvb-eps-p2-peps} by the map
\begin{equation}
        \label{eq:orvb-proj}
\mc P_{\perp} = \ket{02}\bra{02}+\ket{12}\bra{12} + 
                     \ket{20}\bra{20}+\ket{21}\bra{21}\ .
\end{equation}
It is straightforward to check that the resulting state is an equal weight
superposition of all dimer coverings, and that different dimer
configurations are orthogonal, since the position of the dimers can be
unambiguously inferred from the location of the $\ket{2}$'s. Note that
while we write the physical space as a subspace of a $3\times
3$--dimensional space, in fact only a four-dimensional subspace is used.
A more detailed discussion of the construction can again be found in
Appendix~\ref{appendix:rvb-peps-representations}.

As one might expect, it is possible to obtain the RVB state from the dimer
state by coherently discarding the information about where the singlet
is located and only keeping the singlet subspace
$\mathrm{span}\{\ket{0},\ket1\}$. This is most easily seen by noting that
\[
\mc P= \mc P \mc P_\perp\ ,
\]
and thus, applying $\mc P$ to each site of the dimer state yields the RVB
state. 

Note that our representation of the dimer state differs from the
formulations typically used in the literature in two points:  Firstly,
while dimer models have been usually studied on the space spanned by valid
dimer configurations, our formulation allows to locally check whether a
dimer is present on a given link.  This allows us to enforce the
restriction to the space of valid dimer configurations using a local
Hamiltonian, and ultimately enables us to map our dimer model to the toric
code by local unitaries. Secondly, unlike mappings where the presence or
absence of a dimer is represented by a spin,~\cite{nayak:dimer-models} our
mapping uses actual singlets built of two entangled particles.  This
allows us to locally interpolate between the dimer and the RVB model, and
gives rise to a gapped entanglement spectrum, which is in contrast to the
gapless entanglement spectra which had been observed previously for
topologically ordered systems.

\subsection{Gauge transformation}

The PEPS representation of the RVB and dimer state are built using unnormalized
singlets, as the weight of $\ket{01}-\ket{10}$ and $\ket{22}$ in
(\ref{eq:eps-bond}) is the same.  Since the number of singlets in all
dimer configurations is the same, this is not an issue; however, in some
situations (cf.~Sec.~\ref{sec:tc-dimer-mapping}) it
will be more convenient to normalize the singlets, i.e., replace
(\ref{eq:eps-bond}) by 
\begin{equation}
\label{eq:eps-bond-gauge}
\ket{\hat\varepsilon} = 
    \tfrac{1}{\sqrt{2}}\sum_{i,j,k=0}^2 \varepsilon_{ijk}\ket{ijk} + \ket{222}\ .
\end{equation}
Note that $\ket{\hat\varepsilon} = (Y\otimes Y\otimes
Y)\ket{\varepsilon}$, where $Y=\mathrm{diag}(2^{-1/4},2^{-1/4},1)$.  Since
on the other hand, $\mc P_\perp (Y\otimes Y) = 2^{-1/4}\mc P_\perp$ (and thus also
for $\mc P$, as well as the interpolation defined in
Sec.~\ref{sec:interpolation}), we find that these two PEPS do not only
represent the same total state, but are in fact related by a \emph{local}
gauge transformation, allowing to directly relate all properties of their
PEPS description.

\subsection{The toric code}

\begin{figure}[b]
\includegraphics[width=\columnwidth]{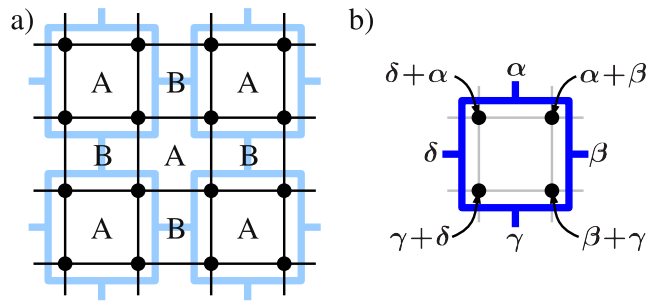}
\caption{\label{fig:toric-code}
Construction of the toric code as a tensor network. \textbf{a)} The toric
code is defined on a square lattice with two types of plaquettes. By
taking $2\times 2$ blocks over $A$--type plaquettes, a particularly
convenient tensor network description can be found. \textbf{b)} Tensor for
the toric code; depicted are the non-zero entries. The four auxiliary
indices are in the $X$ basis, while the physical sites are in the $Z$
basis.}
\end{figure}

Consider a square lattice of qubits, with an $A$-$B$-pattern on the
plaquettes, Fig.~\ref{fig:toric-code}a. Kitaev's toric
code state~\cite{kitaev:toriccode} is the ground state of the Hamiltonian
\begin{equation}
        \label{eq:tc-ham}
H_{\mathrm{TC}}=\sum_{p_A}\tfrac12(1-{Z}^{\otimes 4}_{p_A})
    + \sum_{p_B}\tfrac12(1-{X}^{\otimes 4}_{p_B})\ ,
\end{equation}
where the sums run over all $A$ and $B$ type plaquettes $p_A$ and $p_B$,
respectively, and the tensor products of Paulis act on the qubits adjacent
to each plaquette.  

The toric code state can be written as a
PEPS~\cite{verstraete:comp-power-of-peps} with bond dimension $D=2$. A
particularly useful representation is obtained by blocking the qubits in
$2\times 2$ blocks around $A$ type plaquettes, as indicated in
Fig.~\ref{fig:toric-code}a and letting the PEPS projector
be~\cite{schuch:peps-sym}
\begin{equation}
        \label{eq:tc-proj}
\mc P_{\mathrm{TC}} = \sum_{\alpha,\beta,\gamma,\delta=0}^1
    \ket{\alpha+\beta,\beta+\gamma,\gamma+\delta,\delta+\alpha}
    \bra{\hat\alpha,\hat\beta,\hat\gamma,\hat\delta}\ ,
\end{equation}
for each of these blocks. Here, the sums are modulo $2$, and the ordering
of the indices is illustrated in Fig.~\ref{fig:toric-code}b; the virtual
qubits are expressed in the $X$ eigenbasis $\ket{\hat
0}=(\ket0+\ket1)/\sqrt{2}$, $\ket{\hat
1}=(\ket0-\ket1)\sqrt{2}$.

\subsection{Formulation using tensor networks
\label{sec:tn-formulation-of-rvb}}

PEPS can also be expressed in terms of
\emph{tensor networks}; while both descriptions are mathematically
equivalent we will utilize either of them when more convenient.

\begin{figure}
\includegraphics[width=0.9\columnwidth]{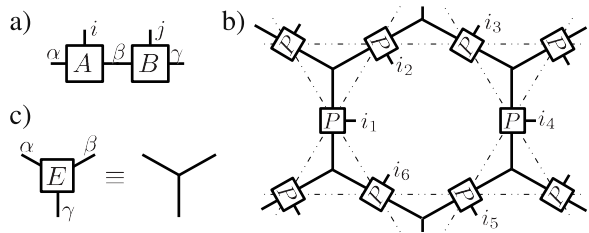}
\caption{\label{fig:rvb-tn-description}
Tensor network notation for the RVB state. \textbf{a)} Tensors are denoted
as boxes with one leg per index.  Connecting legs corresponds to summing
over that index, panel a) e.g.\ shows $\sum_\beta
A^i_{\alpha\beta}B^j_{\beta\gamma}$.
\textbf{b)} Tensor network description of the RVB state; the triangluar
link is shorthand for the tensor $E$ implicitly defined in
Eq.~(\ref{eq:E-tens-def}), see  panel c). Depending on whether we
choose $P$ as in Eq.~(\ref{eq:P-tensor}) or replace it by $P_\perp$, 
Eq.~(\ref{eq:Pperp-tensor}), we obtain a tensor network for either the RVB
or the dimer state; in the latter case, the physical indices $i_k$
form double indices $(i_k,j_k)$.}
\end{figure}

A tensor network state is a state
\begin{equation}
        \label{eq:state-w-coefficient-tn}
\ket{\psi} = 
\sum_{i_1,\dots,i_N=0}^1 c_{i_1\dots i_N}\ket{i_1,\dots,i_N}\ ,
\end{equation}
where $c_{i_1\dots i_N}$ can be expressed efficiently by a tensor network,
i.e., as the sum over $\alpha_1,\dots,\alpha_M$ of product of tensors with
indices $i_1,\dots, i_N$ and $\alpha_1,\dots,\alpha_M$,  where each of the
tensors only has a few indices.  Tensor networks are often expressed
graphically, with tensors denoted by boxes with legs corresponding to the
indices, where connecting legs corresponds to contracting over the
corresponding index, cf.~Fig.~\ref{fig:rvb-tn-description}a.

It is straightforward to see that the simplified PEPS representation of
the RVB state can be expressed using two types tensors, $E$ and $P$,
defined such that
\begin{equation}
\label{eq:E-tens-def}
\ket\varepsilon \equiv
\sum_{\alpha\beta\gamma}E^{\alpha\beta\gamma}\ket{\alpha,\beta,\gamma}
\end{equation}
and
\begin{equation}
\label{eq:P-tensor}
\mc P \equiv \sum_{i\eta\vartheta} P^i_{\eta\vartheta}\ket{i}\bra{\eta,\vartheta}
\ ,
\end{equation}
which form the tensor network of Fig.~\ref{fig:rvb-tn-description}b (where
we use the shorthand Fig.~\ref{fig:rvb-tn-description}c for $E$); it has
open indices $i_1,\dots,i_N$ and yields the coefficient
$c_{i_1,\dots,i_N}$ in Eq.~(\ref{eq:state-w-coefficient-tn}). Again, we
can replace $P$ by $P_\perp$ defined by
\begin{equation}
\label{eq:Pperp-tensor}
\mc P_{\perp} \equiv \sum\limits_{ij\eta\vartheta} 
(P_\perp)^{ij}_{\eta\vartheta}\ket{i,j}\bra{\eta,\vartheta}
\end{equation}
to obtain a
PEPS representation of the dimer state.

\section{Relations
\label{sec:relations}}

In this section, we will show how one can transform 
reversibly between the toric
code, the dimer state, and the RVB state. To this end, we will start by
recalling the concept of $G$-injectivity of PEPS~\cite{schuch:peps-sym}
and subsequently apply it it to prove equivalence of the toric code and
the dimer state, and to devise a mapping which transforms between the
dimer and the RVB state.

\subsection{$\mathbb Z_2$--injectivity
\label{sec:z2-injectivity}}

\begin{figure}
\includegraphics[width=\columnwidth]{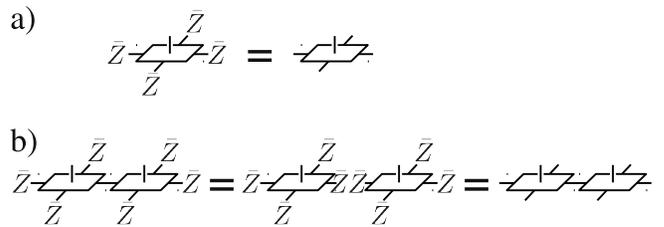}
\caption{\label{fig:z-invariant-tn}
$\mb Z_2$--invariant tensor network. \textbf{a)} A tensor is called $\mb
Z_2$--invariant if it is invariant unter some $\bar Z$ applied to all
virtual indices, where $\{1,\bar Z\}$ forms a representation of $\mathbb
Z_2$.  \textbf{b)} $\mb Z_2$--invariance is stable under concatenation of
tensors.}
\end{figure}

We start by introducing \emph{$\mb Z_2$--injectivity}, which has been
introduced (as $G$--injectivity, with $G$ any finite group) in
Ref.~\onlinecite{schuch:peps-sym} as a tool to characterize PEPS with
topological properties.  
Let $\bar Z$ be a unitary such that $\{\openone,\bar Z\}$ forms a
representation of $\mathbb Z_2$, i.e., $\bar Z^2=\openone$. A
\emph{$\mathbb Z_2$--invariant} tensor network consists of tensors 
which are invariant under the symmetry action on all virtual indices
simulateously, as depicted in Fig.~\ref{fig:z-invariant-tn}a (we restrict
to equal representations on all indices).  This property 
is stable under blocking, i.e., two (or more) $\mb Z_2$--invariant tensors
together are still $\mb Z_2$--invariant, see Fig.~\ref{fig:z-invariant-tn}b.
A $\mb Z_2$--invariant tensor is called \emph{$\mb Z_2$--injective} if the
corresponding PEPS projector $\mc P$ is injective on the invariant
subspace (the subspace belonging to the trivial representation),
i.e., if it has a left inverse $\mc P^{-1}$ such that $\mc P^{-1}\mc P$ is
the projection onto the invariant subspace.  Differently
speaking, this means that there exists a linear map which we can apply to
the physical system to obtain direct access to the virtual system, up
to a projection onto the symmetric subspace $\Pi_S$. $\mb
Z_2$--injectivity is stable under concatenation: When composing two $\mb
Z_2$--injective tensors, we obtain another $\mb Z_2$--injective
tensor.~\cite{schuch:peps-sym}

The key point to note is that any two $\mb Z_2$--injective tensor networks
with identical geometry and symmetry representation, but different
tensors, are locally equivalent:  There exists a map acting on
the physical system of individual tensors which transforms the networks
into another. This follows since both PEPS projectors, $\mc P_A$ and
$\mc P_B$, have left inverses which yield the projector $\Pi_S$ onto the
symmetric subspace, $\mc P^{-1}_A\mc P_A=\mc P^{-1}_B\mc P_B=\Pi_S$, and
thus, $\mc P_A = (\mc P_A \mc P^{-1}_B)\mc P_B$, and vice versa.

An important special case are \emph{$\mathbb Z_2$--isometric} tensors
where $\mc P$ is a partial isometry, $\mc P^{-1}=\mc P^\dagger$: Any two
$\mathbb Z_2$--isometric tensors $\mc P_A$ and $\mc P_B$ can be
transformed into each other by a unitary acting on the physical system.
Again, $\mathbb Z_2$--isometry is a property which is stable under
concatenation.~\cite{schuch:peps-sym}

\subsection{\label{sec:tc-dimer-mapping}
Toric code and dimer state}

In the following, we will show that the Toric Code state and the
orthogonal dimer state are both built of $\mathbb Z_2$--isometric tensors,
which implies that they can be transformed into each other by local
unitaries. (This relation has been observed previously by mapping the
presence or absence of dimers to a spin degree of
freedom.\cite{nayak:dimer-models}) In this section, we will
consider the
$\mathbb Z_2$ representation $\{\openone,Z\}$, with
$Z=\left(\begin{smallmatrix}1\\&-1\end{smallmatrix}\right)$.

\begin{figure}
\includegraphics[width=\columnwidth]{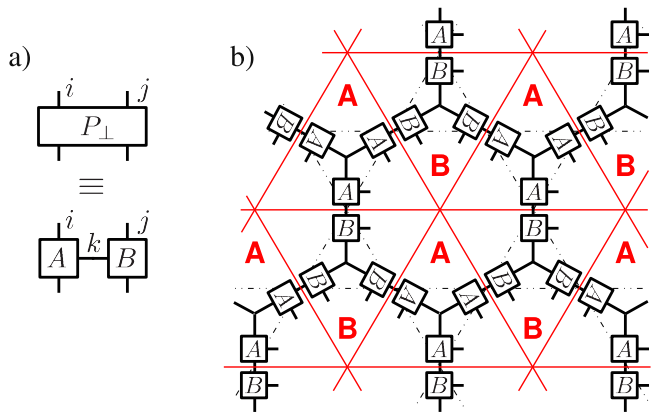}
\caption{\label{fig:split-pperp-triangles}
Transformation of the dimer state into the toric code.
\textbf{a)} $P_\perp$ can be rewritten as a tensor network with bond
dimension $2$. \textbf{b)} Network obtained by replacing $P_\perp$ in
Fig.~\ref{fig:rvb-tn-description}b by the tensor network of panel a). The
tensor network can be grouped into triangular tensors which contain only
$A$ or $B$ tensors, and which are connected by bonds of dimension $2$.
}
\end{figure}

For the toric code tensor Eq.~(\ref{eq:tc-proj}), it is straightforward to
check that it is invariant under $Z^{\otimes 4}$, and acts isometrically on
the invariant subspace, i.e., is $\mathbb Z_2$--isometric.  In order to
see the same for the dimer state, we first rewrite
$P_\perp$, Eq.~(\ref{eq:Pperp-tensor}), as
\begin{equation}
\label{eq:p-perp-tn} P_{\perp} = \sum_{k=0}^1 A_k\otimes B_k\ ,
\end{equation}
where $A_0 = B_1 = \ketbra{0}+\ketbra1$, and $A_1 = B_0 = \ketbra2$. Using
this decomposition, we replace each tensor $P_\perp$ in
Fig.~\ref{fig:rvb-tn-description}b with a pair of tensors $A$ and $B$
connected by a two-dimensional bond,
cf.~Fig.~\ref{fig:split-pperp-triangles}a, oriented as
in Fig.~\ref{fig:split-pperp-triangles}b. This orientation allows to block
the tensors into triangles, labelled $A$ and $B$, containing only $A$ and
$B$ type tensors, cf.~Fig.~\ref{fig:split-pperp-triangles}b; the
triangles are connected
by bonds of dimension two.

\begin{figure}
\includegraphics[width=\columnwidth]{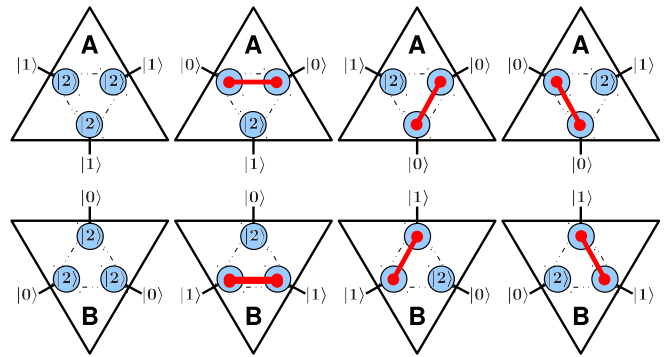}
\caption{
\label{fig:triangle-configs}
Non-zero orthogonal virtual states and the corresponding orthogonal
physical states for the dimer state tensors (the red lines denotes
singlets). The $A$ ($B$) type tensors are supported on the odd (even)
parity subspace.
}
\end{figure}

The bonds are associated to vertices; a bond state $\ket{0}$ ($\ket{1}$)
indicates that the dimer lies inside the $A$ ($B$) triangle (similar to
the ``arrow representation'' of
Ref.~\onlinecite{elser:rvb-arrow-representation}).  As each triangle
contains either no or one dimer, the $A$ ($B$) tensors are odd (even)
under $Z^{\otimes 3}$ symmetry, and moreover isometric on the
(anti)symmetric subspace if we use the gauge of Eq.~(\ref{eq:eps-bond-gauge});
the possible mappings from bond to physical configurations are given in
Fig.~\ref{fig:triangle-configs}.

We now group one $A$ and one $B$ tensor to obtain square tensors, which
are antisymmetric and isometric; grouping two of these square tensors
finally yields a $\mathbb Z_2$--isometric tensor with symmetry
representations $Z$ and $Z\otimes Z$, respectively.  Blocking two tensors
of the toric code also yields a $\mathbb Z_2$--isometric tensor with the
very same symmetry, and thus, the dimer state and the toric code can be
converted into one another by local unitaries.

\subsection{RVB and dimer state}

Let us now show that we can reversibly (though not unitarily) transform
the dimer state to the RVB state. In principle, the forward direction is
obvious, as $\mc P = \mc P\mc P_\perp$; however, this transformation
cannot be inverted. In order to construct an invertible transformation, we
will show that both states can be understood as being constructed from
$\mathbb Z_2$--injective tensors connected by $\ket\varepsilon$ bonds:
Following the reasoning of Sec.~\ref{sec:z2-injectivity}, we can
reversibly transform between the two $\mathbb Z_2$--injective tensors and
thus between the RVB state and the dimer state.

\begin{figure}[t]
\begin{center}
\includegraphics[width=.95\columnwidth]{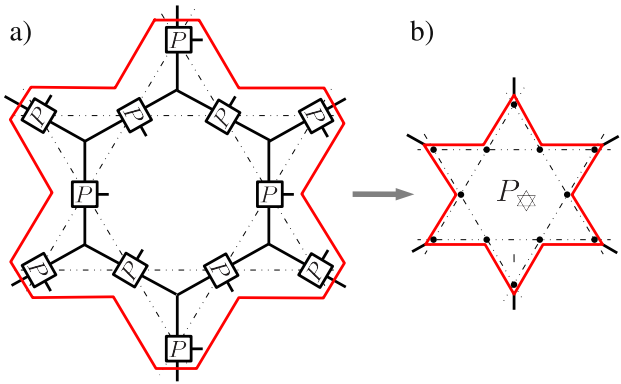}
\end{center}
\caption{\label{fig:pstar-definition}
Definition of the tensor $P_{\ds}$.  \textbf{a)} $P_{\ds}$ is obtained by
blocking $12$ tensors $P$ with $6$ triangular ``bond tensors'' $E$; it
maps $6$ virtual qutrits to $12$ physical qubits. \textbf{b)}~Effective
depiction of the tensor $P_{\ds}$.
}
\end{figure}

\begin{figure}[t]
\includegraphics[width=0.9\columnwidth]{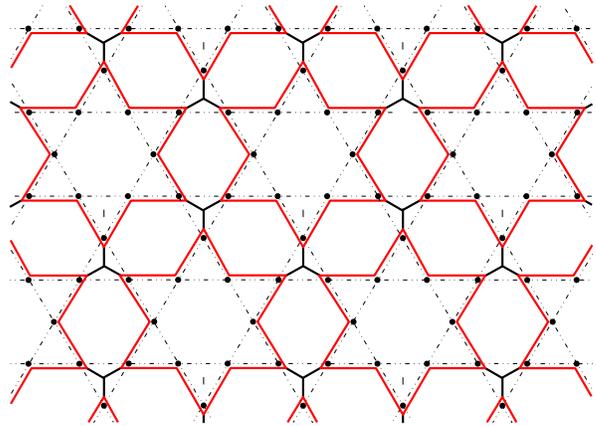}
\caption{\label{fig:star-covering}
Complete covering of the kagome lattice with stars,
Fig.~\ref{fig:pstar-definition}. The stars tensor $P_{\ds}$ (or projectors
$\mc P_{\ds}$),
together with triangular tensors $E$ (or $\ket\varepsilon$ bonds),
describe the RVB state (or its orthogonal version, if we choose 
$P_{\perp,\ds}$ instead).  By acting with $\mc O_{\ds}$ on each of the
stars, we can reversibly convert between the RVB and the orthogonal RVB.}
\end{figure}

For this section, we will use the PEPS representation
Fig.~\ref{fig:rvb-tn-description} of both the RVB and the dimer state.
The representation of the $\mb Z_2$ symmetry is given by 
\[
\ZZ := \left(\begin{matrix}
            1\\&1\\&&-1
    \end{matrix}\right)\ .
\]
It is straightforward to see that both $P$, $P_{\perp}$, and
the bond tensor $E$ are anti-symmetric under the action of the
symmetry on all virtual indices.
Let us now form a new tensor by blocking all the tensors covering one star,
cf.~Fig.~\ref{fig:pstar-definition}. We call the new tensors $P_{\ds}$
and $P_{\perp,\ds}$, respectively (with corresponding PEPS projector $\mc
P_{\ds}$ and $\mc P_{\perp,\ds}$). Clearly, we can tile the lattice with
$P_{\ds}$ tensors connected by $E$ bonds, see
Fig.~\ref{fig:star-covering}.  Due to stability of $\mb Z_2$--invariance,
one finds that $P_{\ds}$ and $P_{\perp,\ds}$ are $\mb Z_2$--invariant.  On
the other hand, it can be checked analytically using a computer algebra
system that $\mc P_{\ds}$ is indeed invertible on the invariant subspace
of $\ZZ^{\otimes 6}$, i.e., $\mc P_{\ds}$ is $\mathbb Z_2$--injective. 
This immediately implies $\mathbb Z_2$--injectivity of $P_{\perp,\ds}$ (as
$\mc P_{\ds}=\mc P^{\otimes 6}\mc P_{\perp,\ds}$).

We can now use the argument of Sec.~\ref{sec:z2-injectivity} to construct
an invertible mapping: $\mb Z_2$--injectivity of $\mc P_{\ds}$ implies
the existence of $\mc P^{-1}_{\ds}$ such that $\mc P^{-1}_{\ds}\mc
P_{\ds}$ is the projector onto the $\ZZ^{\otimes 6}$--invariant subspace
on the virtual qutrits along the boundary of the star. If we now define $\mc
O_{\ds}=\mc P_{\perp,\ds}\mc P_{\ds}^{-1}$, it follows that $\mc
O_{\ds}\mc P_{\ds} = \mc P_{\perp,\ds}$, i.e., it (locally) converts the
RVB into the dimer state.  [Note that $\mc O_{\ds}$ is only well-defined on
the range of $\mc P_{\ds}$, $\rg(\mc P_{\ds})$; we choose $\mc O_{\ds}$ to
vanish on its orthogonal complement.] Analogously, we can define $\mc
O^{-1}_{\ds}$ such that $\mc O^{-1}_{\ds}\mc P_{\perp,\ds} = \mc P_{\ds}$
[again such that it vanishes on $\rg(\mc P_{\perp,\ds})^{\perp}$], and it
is easily verified that $\mc O^{-1}_{\ds}\mc O_{\ds} = \openone_{\rg(\mc
P_{\ds})}$ and $\mc O_{\ds}\mc O^{-1}_{\ds} = \openone_{\rg(\mc
P_{\perp,\ds})}$, i.e., $O_{\ds}^{-1}$ is the pseudoinverse of $O_{\ds}$.

Applying the maps $\mc O_{\ds}$ or its inverse to all $K=N/12$ stars in the
tiling Fig.~\ref{fig:star-covering}, we can now reversibly convert between
the dimer and the RVB state, $\ket{\Psi_{\mathrm{dimer}}}= \mc
O_{\ds}^{\otimes K}\ket{\Psi_{\mathrm{RVB}}}$ and
$\ket{\Psi_{\mathrm{RVB}}}=\mc (O^{-1}_{\ds})^{\otimes
K}\ket{\Psi_{\mathrm{oRVB}}}$.

\section{Interpolation
\label{sec:interpolation}}

\subsection{Interpolation between dimer model and RVB
\label{sec:dimer-rvb-interpol}}

Let us now show how to continuously deform the dimer state into the RVB
state; as we will show in Sec.~\ref{sec:ham-dimer-rvb-interpol}, this
interpolation does correspond to a continuous change of the associated 
local parent
Hamiltonian. To this end, we define a family of PEPS projectors
\begin{equation}
        \label{eq:P-interpolation}
\begin{aligned}
\mc P(\theta) &= 
\ket{+}\Big[\ket{0}(\bra{02}+\bra{20})+\ket{1}(\bra{12}+\bra{21})\Big]
\\
& +\theta\,\ket{-}
    \Big[\ket{0}(\bra{02}-\bra{20})+\ket{1}(\bra{12}-\bra{21})\Big]
\end{aligned}
\end{equation}
with a two-qubit physical space.
Clearly, $\mc P(0) \equiv \mc P$,
and $\mc P(1)\equiv \mc P_\perp$, both up to local isometries. By
placing $\mc P(\theta)$ in the tensor network
Fig.~\ref{fig:rvb-eps-p2-peps}, we therefore obtain a smooth interpolation
$\ket{\Psi(\theta)}$
between the RVB and the dimer state. Clearly, we can accordingly define 
$P_{\ds}(\theta)$ and subsequently $\mc O_{\ds}(\theta) = \mc
P_{\ds}(\theta) \mc P_{\ds}^{-1}$, which is  a continuous invertible map 
from $\rg\,\mc P_{\ds}$ to $\rg\,\mc P_{\ds}(\theta)$.
Note that while it might seem that the interpolation (\ref{eq:P-interpolation})
results in a discontinuity as $\theta\rightarrow0$, the discussion about
the equivalence of RVB and dimer state in the preceding section shows that
this discontinuity disappears when considering whole stars.

This interpolation can be understood as a way to make different dimer
configurations more and more orthogonal, analogous to what is achieved by
decorating the lattice in Ref.~\onlinecite{raman:rvb-decorated-lattice}:
Overlapping two different dimer configurations gives rise to a loop
pattern, and the absolute value of their overlap is $x^{L-2N}$, with $N$
the number of loops (including length-$2$ ones) and $L$ their total length
(which equals the number of lattice sites).  Here, $x=1/\sqrt{2}$ for the
RVB and $x=0$ for the dimer state, and Eq.~(\ref{eq:P-interpolation})
corresponds to an interpolation
$x(\theta)=\tfrac{1}{\sqrt{2}}\tfrac{1-\theta^2}{1+\theta^2}$.

\subsection{Numerical results}

\begin{figure}[t]
\includegraphics[width=\columnwidth]{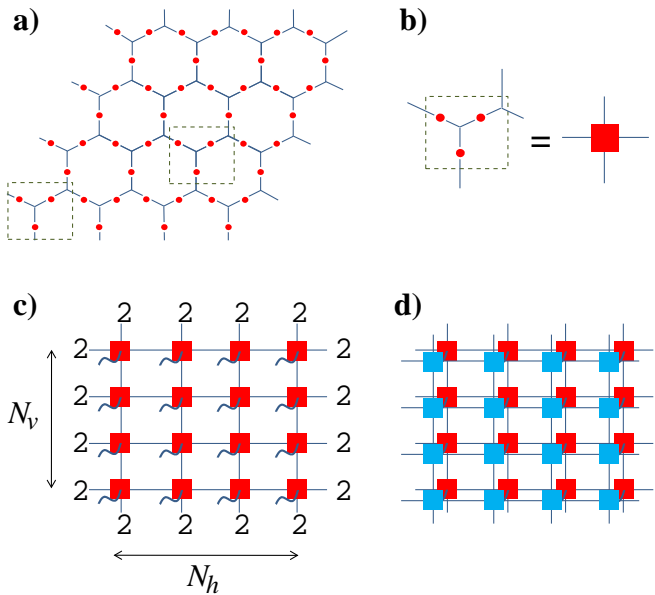}
\caption{
\label{fig:num-lattice}
\textbf{a)} Lattice used for the numerical calculations. The red dots
denote $P(\theta)$ tensors. \textbf{b)} By blocking three spins, we
transform this to a square lattice.  \textbf{c)} Resulting square lattice.
For OBC, we set the bonds at the boundary to $\ket{2}$ as shown; for CBC,
we consider a vertical cylinder where left and right indices are
connected. \textbf{d)} Tensor representation of the scalar product
$\langle \Psi(\theta)|\Psi(\theta')\rangle$.  }
\end{figure}

\begin{figure}[b]
\includegraphics[width=0.8\columnwidth]{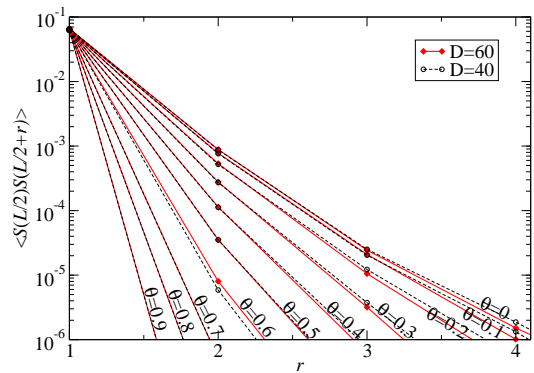}
\caption{\label{fig:num-corr-p-p}
Two-point correlations as a function of the distance for different values
of $\theta$. The plot shows data for an infinite system where in the PEPS
contraction,\cite{verstraete:2D-dmrg} the bond dimension has been
truncated at $D=40$ and $D=60$, respectively.}
\end{figure}

We have numerically studied the interpolation (\ref{eq:P-interpolation})
between the dimer state (which is a topological $\mathbb Z_2$ spin liquid)
and the RVB state to determine whether they are in the same phase.  We use
the tensor network representation of Fig.~\ref{fig:rvb-tn-description},
where the interpolating path is characterized by tensors $P(\theta)$
corresponding to $\mc P(\theta)$ of Eq.~(\ref{eq:P-interpolation}).  For
the numerical calculations, it is particularly convenient to block three
spins, i.e., $P(\theta)$ tensors, together with two $E$ tensors, as
indicated by the dashed squares in Fig.~\ref{fig:num-lattice}a; this
allows us to rewrite the system on a square lattice with three spins per
site (Fig.~\ref{fig:num-lattice}b).  We perform simulations both for open
boundary conditions and for cylindrical boundary conditions.
For open boundary conditions (OBC), we set all open indices in the tensor
network to $\ket2$ (i.e., there are no outgoing singlets at the edges), as
indicated in Fig.~\ref{fig:num-lattice}c. The case of cylindrical
boundary conditions is obtained by putting the system on a vertical
cylinder, i.e., connecting the left and right indices, while the upper and
lower outgoing indices are still set to $\ket{2}$.  We will denote by
$N_h$ and $N_v$ the number of sites in each row and column of the square
lattice, respectively.   We will generally consider the case where
$N_v\rightarrow\infty$ for finite $N_h$. In this case, cylindrical
boundary conditions give the same results as periodic boundary conditions
as long as topological degeneracies are appropriately taken into account,
and we will therefore refer to them as periodic boundary conditions (PBC)
in the
following.

\subsubsection{Correlation functions and gap of the transfer operator}

\begin{figure}[t]
\includegraphics[width=0.9\columnwidth]{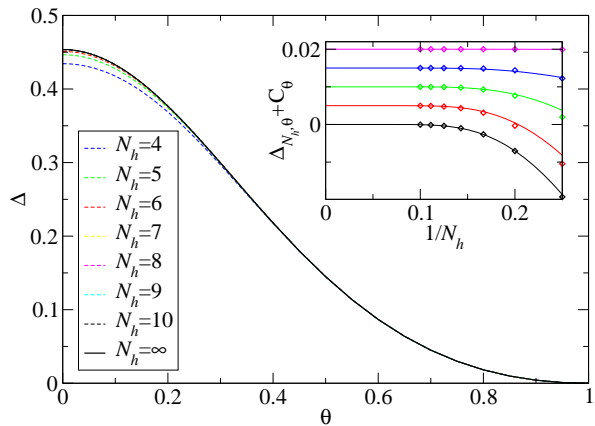}
\caption{\label{fig:num-top-gap}
Gap of the transfer operator $\Delta$ as a function of $\theta$ along the
RVB--dimer interpolation for $N_h=4,\dots,10$ and PBC.
As the transfer operator is parity preserving (changing)
for $N_h$ even (odd), it has two (almost) degenerate largest eigenvalues
corresponding to different parity sectors (for $N_h$ even, there is an
exponentially small splitting in $N_h$);  the gap $\Delta$ shown is
the ratio of the third eigenvalue and the average of the two maximal
eigenvalues.
The data for $N_h=\infty$ has been obtained by fitting the data
for $N_h=6,\dots,10$ with \mbox{$a+b\,\exp(-cN_h)$}.  The inset shows the fit for
$\theta=0,0.1,0.2,0.3,0.5$ (from bottom to top, the curves have been
shifted vertically in the plot); note that the two rightmost data points
have not been used for the fit.
} 
\end{figure}

Firstly, we have studied the decay of correlation functions along the
dimer--RVB interpolation.  Correlation functions can be computed for
lattices of arbitrary size using standard methods for the contraction of
PEPS.\cite{verstraete:2D-dmrg} We have computed various correlation
functions, all of which show an exponential decay with a slope which
changes smoothly with $\theta$, and do not exhibit any sign of a phase
transition; as an example, Fig.~\ref{fig:num-corr-p-p} shows the connected
dimer-dimer correlation function between two $z\otimes z$ operators
[$z=\mathrm{diag}(1,-1,1)$],  where each dimer is supported on one square
tensor of Fig.~\ref{fig:num-lattice}b and the two squares are in the same
row, as a function of the distance $r$ between the squares, for different
values of $\theta$, evaluated on an infinite system.  (This data has been
obtained by row-wise contraction using infinite Matrix Product States and
iTEBD, and then evaluating the correlation function using the fixed point
tensor.)

The decay of any correlation function is bounded by the gap of the
transfer operator, i.e., one row in Fig.~\ref{fig:num-lattice}d.  We
have used exact diagonalization to determine the gap of the transfer
operator for PBC.  The result for $N_h=4,\dots,10$, as well as the
extrapolated data for $N_h=\infty$, is shown in Fig.~\ref{fig:num-top-gap}.
Again, the gap of the transfer operator stays finite and converges rapidly
for growing $N_h$.  Note that the fact that the ground state is a PEPS
throughout the interpolation (and thus satisfies an area law) does not
preclude critical
behavior.~\cite{verstraete:comp-power-of-peps,barthel:peps-mera}

\subsubsection{Overlap}

In order to detect phase transitions which are not reflected in
correlation functions, we have also studied the rate at which the state
changes as we change $\theta$, as quantified by the overlap (or
fidelity).~\cite{zanardi:overlap-criterion} Concretely, let
$|\Psi(\theta)\rangle$ be the state interpolating between
$|\Psi_{\rm RVB}\rangle\equiv|\Psi(0)\rangle$ and $|\Psi_{\rm
dimer}\rangle\equiv|\Psi(1)\rangle$, as defined in the preceding section.
We define the change of overlap
\[ 
F_{N_h,N_v}(\theta,\epsilon) := 
    \frac{1}{\epsilon^2}
    \left[ 1- \frac{
    |\langle \Psi(\theta+\epsilon)| \Psi (\theta)\rangle|^2}{
    \langle \Psi(\theta+\epsilon)|\Psi(\theta+\epsilon)\rangle \langle
\Psi(\theta)|\Psi(\theta)\rangle} \right] 
\] 
and are interested in the behavior of the limit 
\begin{equation}
f(\theta) = \lim_{N_v,N_h\to\infty}
\lim_{\epsilon\to 0}\, \frac{1}{N_vN_h}\, F_{N_h,N_v}(\theta,\epsilon) 
\label{eq:num-overlap-limit}
\end{equation}
which describes how quickly the state changes in the thermodynamic limit.

\begin{figure}[t]
\includegraphics[width=0.9\columnwidth]{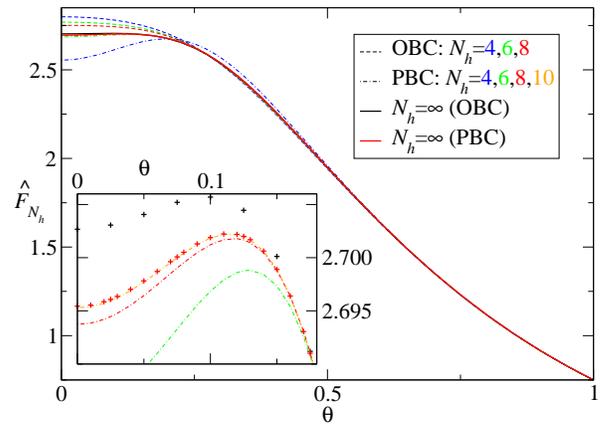}
\caption{
\label{fig:num-overlap-w-inset}
The overlap change $\hat F_{N_h}$, Eq.~(\ref{eq:num-overlap-hatF-def}), as
a function of $\theta$ for both open boundary conditions ($N_h=4,6,8$) and
periodic boundary conditions ($N_h=4,6,8,10$).  The solid black and red
line give the extrapolated data for $N_h=\infty$ for open and periodic
boundaries.  The inset shows a zoom of the $N_h=\infty$ data for
$\theta\le 0.18$, where the black (red) crosses correspond to open
(perodic) boundaries. The details of the scaling analysis are given in
Fig.~\ref{fig:num-overlap-fss}. 
}
\end{figure}

\begin{figure}[b]
\includegraphics[width=0.9\columnwidth]{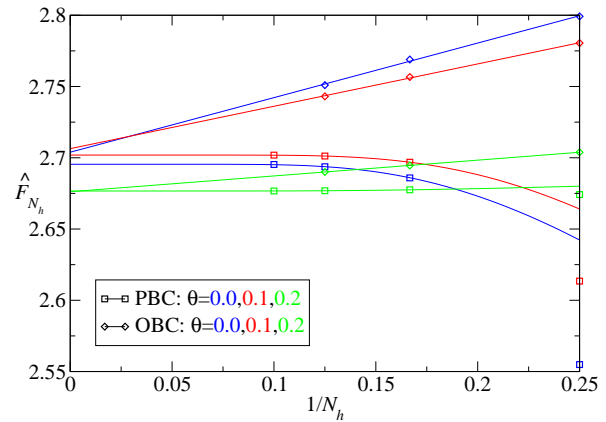}
\caption{
\label{fig:num-overlap-fss}
Finite size scaling analysis for $\hat F_{N_h}$.  The plot shows OBC and PBC
data for $\hat F_{N_h}$. Due to the absence of boundary effects, the PBC
data converges very quickly to the $N_h=\infty$ value once $N_h$ is beyond
the correlation length.  We have fitted the PBC data for $N_h=6,8,10$ with
$a+b\,\exp(-cN_h)$. The OBC
data, on the other hand, has a $1/N_h$ leading term from
the boundary; we have fitted the data for $N_h=4,6,8$ with $a+b/N_h$.  Note
that the correction for the data points for OBC would most likely be
concave, suggesting that the true $N_h=\infty$ value is closer to the PBC
data.  } \end{figure}

\begin{figure*}[t]
\parbox[b]{0em}{\raisebox{18em}{a)}}
\parbox[b]{0.45\textwidth}{
\includegraphics[width=0.45\textwidth]{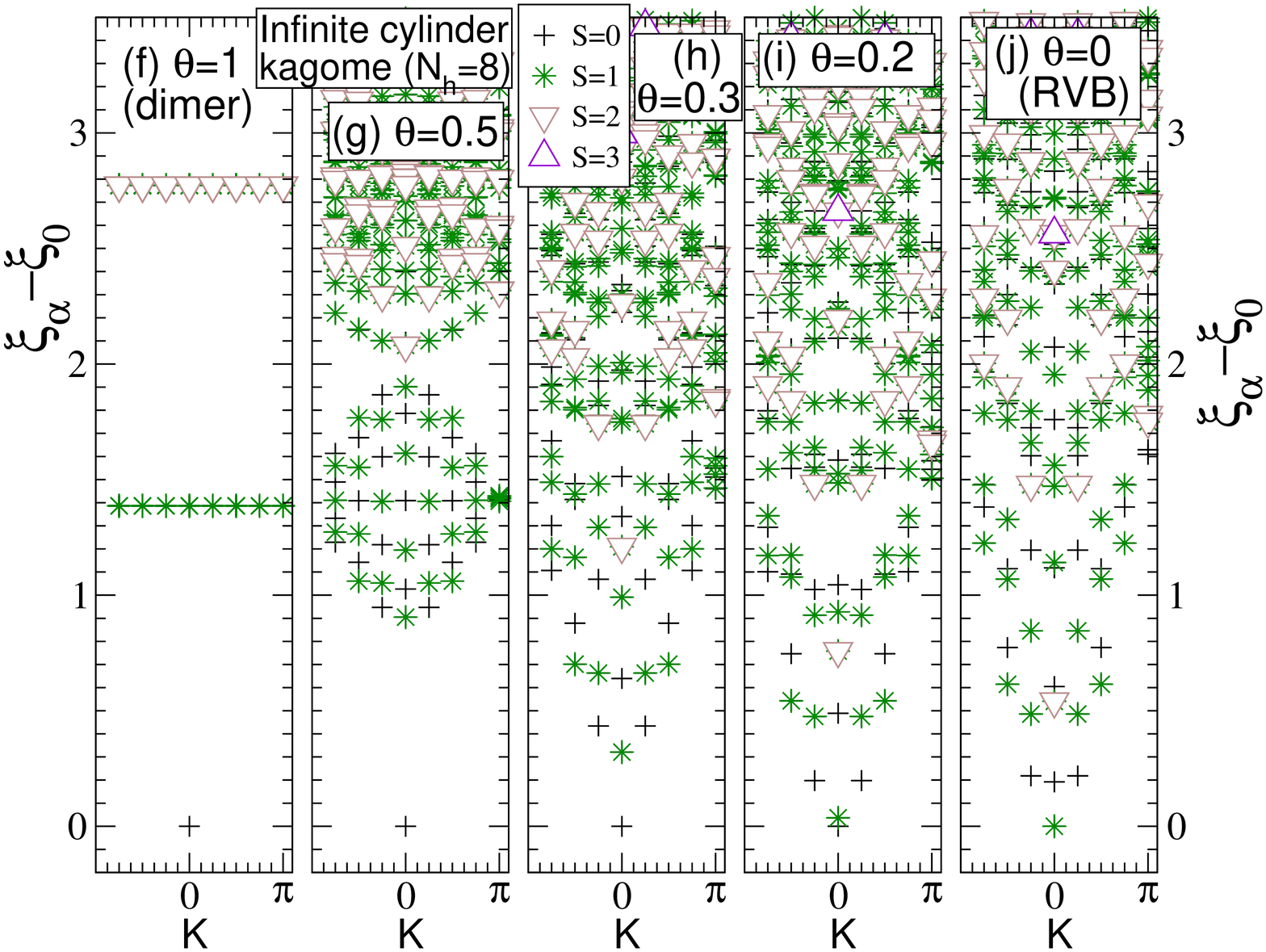}}
\hspace*{1em}
\parbox[b]{0em}{\raisebox{18em}{b)}}
\parbox[b]{0.45\textwidth}{
\includegraphics[width=0.45\textwidth]{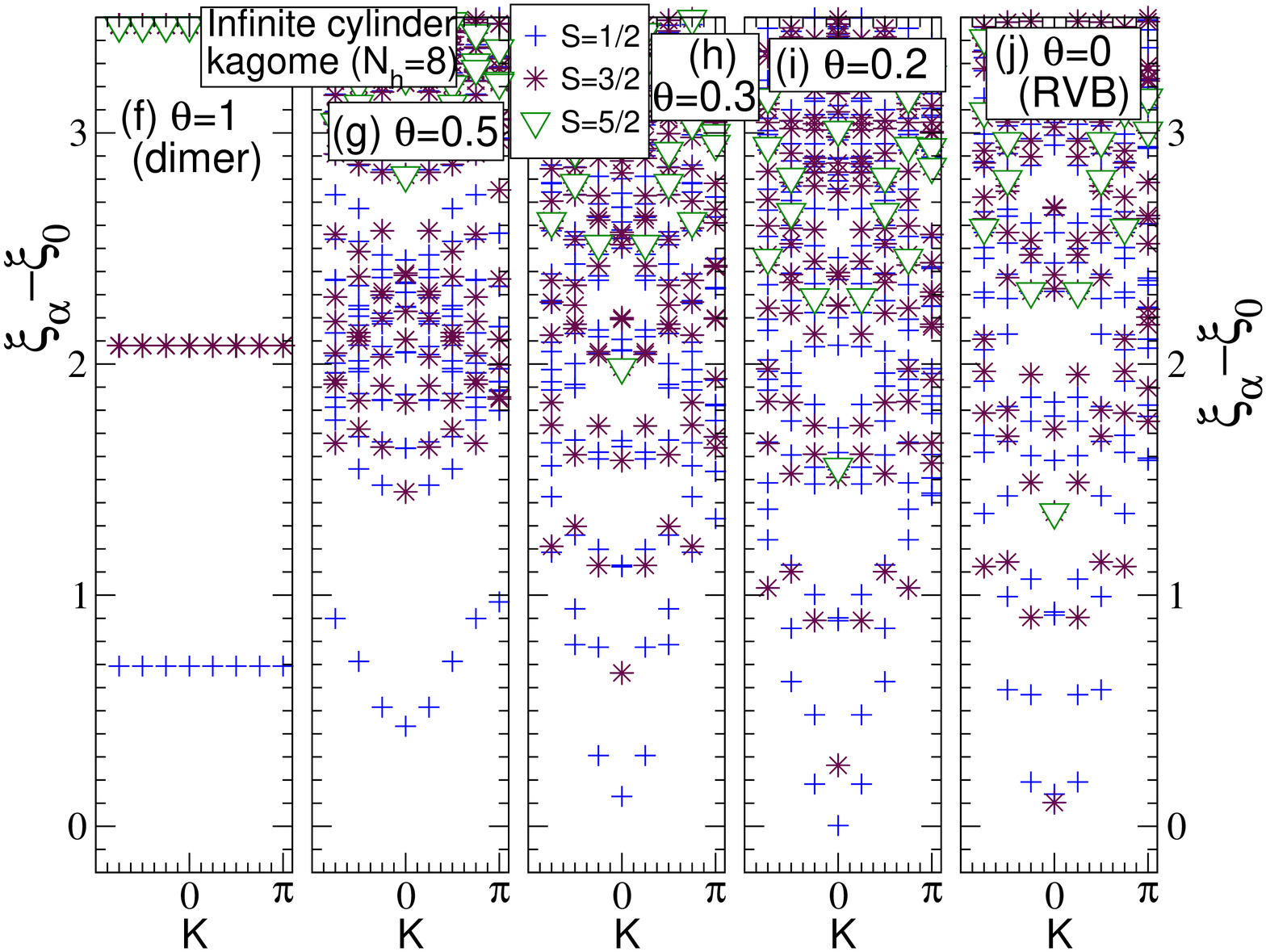}}
\caption{\label{fig:num-entspec}
Entanglement spectrum of an infinite (vertical) cylinder for decreasing $\theta$, and
perimeter $N_h=8$.  The spectrum is shown as a function of the momentum
$K$ along the one-dimensional (horizontal) edge. The eigenstates are also
labelled according to their spin quantum numbers ($\tfrac12\oplus 0$
representation) according to the symbols in the legend.  
Panel a) shows the integer spin sector of the spectrum, panel b) the
half-integer sector; the two sectors are not coupled by the entanglement
Hamiltonian.  Note that for both sectors, the same energy scale has been
chosen (i.e., trace-normalizing the boundary states in both sectors),
corresponding to a local entanglement
Hamiltonian.~\cite{poilblanc:rvb-boundaries}
} \end{figure*}

We have computed this quantity in two different ways. In the first
approach, we use that for sufficiently small $N_h$ we can exactly compute
$F_{N_h,N_v}(\theta,\epsilon)$ for arbitrary values of $N_v$ and
$\epsilon$ by exact row-wise contraction of the tensor network
Fig.~\ref{fig:num-lattice}d, and use this to extrapolate the limit 
\begin{equation}
\label{eq:num-overlap-hatF-def}
\hat F_{N_h}(\theta)=
    \lim_{N_v\rightarrow\infty} 
    \lim_{\epsilon\rightarrow0}
    \frac{1}{N_hN_v}
    F_{N_h,N_v}(\theta,\epsilon)
\end{equation}
to arbitrary accuracy for $N_h=4,6,8$; this data is then used in a finite size scaling
analysis to infer $f(\theta)\equiv \lim \hat F_{N_h}(\theta)$. We apply
this approach to the OBC case.  The second
approach makes use of the fact that $\lim_{\epsilon\rightarrow0}
F_{N_h,N_v}(\theta,\epsilon)$ can be expressed as an average over
correlation functions, as we show in
Appendix~\ref{appendix:overlap-change}.  Again, this average over
correlation functions can be computed exactly for any value of $N_v$, and
$N_h\le10$ (to reach $N=10$, we use that total bond dimension of the
transfer operator of the $P$ tensor, and thus the overall bond dimension
in the tensor network in Fig.~\ref{fig:num-lattice}d, can be reduced from
$D^2=9$ to $6$), which again allows us to determine $\hat F_{N_h}$ for
$N_h=4,6,8,10$, and subsequently apply a finite size scaling analysis.  We
apply this approach to the PBC case.

The resulting data is shown in Fig.~\ref{fig:num-overlap-w-inset},
together with extrapolated data for $N_h=\infty$ from the finite size
scaling. (See Fig.~\ref{fig:num-overlap-fss} for a discussion of the
finite size scaling analysis.) The extrapolated curves for open and
periodic boundary conditions agree very well.  The extrapolated value for
$f(\theta)$ seems to become essentially constant for
$\theta\lesssim0.15$, which might suggest a non-analytic behavior of
$f(\theta)$; however, zooming into this region (inset of
Fig.~\ref{fig:num-overlap-w-inset}) shows clearly that $f(\theta)$
decreases again for $\theta\lesssim 0.12$, showing no evidence
for the presence of a phase transition. (See
Fig.~\ref{fig:num-overlap-fss} for a discussion of the difference observed
between the extrapolated OBC and PBC data.)

\subsubsection{Entanglement spectrum}

\begin{figure}[b]
\includegraphics[width=0.8\columnwidth]{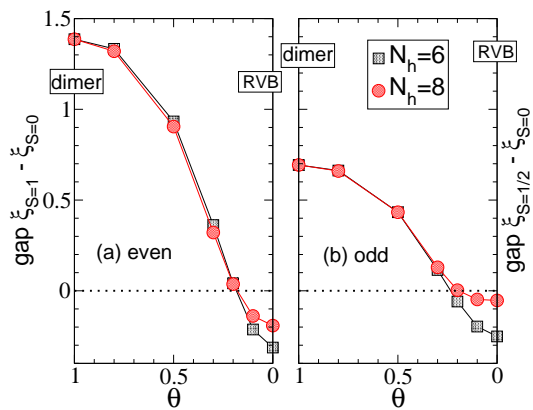}
\caption{\label{fig:num-entgap}
Gap between the lowest $S=1$ and $S=0$ state (left) and $S=\tfrac12$ and
$S=0$ state (right) of the entanglement Hamiltonian for the dimer-RVB
interpolation for $N_h=6,8$.  }
\end{figure}

Finally, we have also studied the behavior of the entanglement spectrum
along the interpolation for an infinite cylinder, using the techniques
described in
Refs.~\onlinecite{cirac:peps-boundaries,poilblanc:rvb-boundaries} (in
particular, using exact contraction).  In
Fig.~\ref{fig:num-entspec} we give the entanglement spectra for $N_h=8$
for the integer and half-integer spin sector (the boundary Hamiltonian
does not couple the two sectors).  While at first, it seems that the
lowest lying $S=\tfrac12$ and $S=1$ states cross with the $S=0$ ground
state at some small $\theta$, a scaling analysis
(Fig.~\ref{fig:num-entgap}) suggests that the gap of the entanglement
Hamiltonian is more likely to vanish at $\theta=0$; also, note that there
is no prior reason to assume that a vanishing gap in the entanglement
spectrum implies any critical behavior in the actual system, as the
boundary state is a thermal state of the entanglement Hamiltonian and
therefore will have finite correlations even for a critical entanglement
Hamiltonian.\cite{cirac:peps-boundaries}

Let us also point out an interesting aspect of the entanglement spectrum:
While the system under consideration is topological (rigorously provable
in an environment of the dimer point~\cite{bravyi:tqo-long}), the
entanglement Hamiltonian has a unique ground state, i.e., it is gapped,
which shows that the connection between topologically ordered systems and
gapless entanglement spectra might be less clear than generally believed.
Note that the reason that the entanglement spectrum is gapped (unlike for
the toric code, the RK model,~\cite{misguich:dimer-kagome} or the
mapping of Ref.~\onlinecite{nayak:dimer-models}) is due to the fact that our
representation of the dimer state involves singlets (i.e., 
an entangled state of two spins) rather than single spins indicating the
presence or absence of a dimer, and that we split these
singlets when computing the entanglement spectrum, just as one would do
for the RVB state itself.

\section{Hamiltonians
\label{sec:hamiltonians}}

\subsection{Hamiltonian for the dimer state}

As has been proven in Ref.~\onlinecite{schuch:peps-sym}, every $G$--injective
PEPS on a square lattice appears as the ground state of a local
Hamiltonian, acting on at most $2\times 2$ elementary cells, which has a
fixed degeneracy determined by the symmetry group.  In
particular, this result can be directly applied to the dimer state to
infer that it is the ground state of a local Hamiltonian with four-fold
degeneracy.  Alternatively,
we can obtain a Hamiltonian for the dimer state by using its local
isometric equivalence to the toric code: The same transformation will also
transform the toric code Hamiltonian into a local Hamiltonian for the
dimer state.  Note, however, that we need to add extra terms to the dimer
Hamiltonian which ensure that the state is constrained to its local
support, as the isometry is only defined as a map between the local
support of the toric code and of the dimer state, respectively.

In Appendix~\ref{appendix:oRVB-parent}, we explicitly derive a Hamiltonian
for the dimer state using the duality with the toric code Hamiltonian.  It
has three types of terms (all of them projectors), acting on vertices,
triangles, and hexagons, respectively:
\begin{itemize}
\item $h_v^\perp$ acts on individual vertices. It ensures that there is
	exactly one singlet per vertex.
\item $h_\medtriangleup^\perp$ acts inside the triangles (as used in the
        $A$--$B$--blocking).  It makes sure that each triangle holds 
        zero or one singlet.
\item $h^\perp_{\hexagon}$ acts on the six vertices adjacent to each hexagon. It
        makes sure that all singlet configurations around the
        corresponding star appear with equal weight. 
\end{itemize}
(The exact form of these three terms, together with their derivation, can
be found in Appendix~\ref{appendix:oRVB-parent}.)

How does this Hamiltonian compare to the Rokhsar-Kivelson (RK) Hamiltonian for
dimer
models?\cite{rokhsar:dimer-models,moessner:dimer-triangular,misguich:dimer-kagome}
The resonance terms $h^\perp_{\hexagon}$ of our Hamiltonian correspond to
those of RK Hamiltonian; in both cases, they ensure that the ground state is the
even weight superposition of all dimer configurations.  On the other hand, our
Hamiltonian has two additional types of terms ($h_v^\perp$ and
$h_\medtriangleup^\perp$) which ensure that the system is in a valid dimer
configuration; these terms are not present in the conventional dimer
models as they are defined right away on the subspace of valid dimer
configurations.  While this makes our Hamltonian more complicated, it is
outweighed by the fact that we have a completely local description of our
system; in particular, it is this locality of the description which allows
us to interpolate between the dimer state and the RVB state locally.
Beyond that, the particular way in which we make the dimer configurations
locally orthogonal allows us to define the resonance moves
$h^\perp_{\hexagon}$ on a single hexagon
(cf.~Appendix~\ref{appendix:oRVB-parent}), as compared to the star
required in the usual RK Hamiltonian.\cite{misguich:dimer-kagome}

\subsection{Hamiltonian for the RVB state}

The Hamiltonian $H^\perp = \sum h_i^\perp$ for the dimer state is
\emph{frustration free}, this is, all Hamiltonian terms---which we have
chosen to be projectors---annihilate the dimer state,
$h_i^\perp\ket{\Psi_\mathrm{dimer}}=0$.  We can make use of this fact to
directly obtain a corresponding frustration free Hamiltonian for
$\ket{\Psi_\mathrm{RVB}}=\mc (O_{\ds}^{-1})^{\otimes K}
\ket{\Psi_\mathrm{dimer}}$:
Define
\begin{equation}
        \label{eq:rvb-ham-from-orvb-ham}
h_i:= \mc O_{\ds}^{\otimes \kappa} 
        h_i^\perp \mc O_{\ds}^{\otimes \kappa}\, 
\end{equation}
where for each $i$, the product $\mc O_{\ds}^{\otimes\kappa}$ only
involves the stars in the covering Fig.~\ref{fig:star-covering} which
overlap with $h_i^\perp$.
Then, $h_i$ is positive semi-definite, and $h_i\ket{\Psi_\mathrm{RVB}}=0$.
Moreover, for each star $k$ we add a term $h_k$ to the Hamiltonian which
projects
onto
the orthogonal complement of $\mathrm{rg}\, \mc P_{\ds}$, and thus
restricts the ground state (=zero-energy) subspace of the overall Hamiltonian to
$\mathrm{rg}\,\mc P_{\ds}^{\otimes K}$. [Note that conversely, those terms
$h_i^\perp$ in $H^\perp$ which restrict the ground state space to
$(\mathrm{rg}\,\mc P^\perp_{\ds})^{\otimes K}$ give $h_i\equiv 0$ in 
Eq.~(\ref{eq:rvb-ham-from-orvb-ham}).] As $\mc O_{\ds}^{\otimes K}$ is a
bijection between $\rg\,\mc P_{\ds}^\perp$ and $\rg\,\mc P_{\ds}$, it
follows that there is a one-to-one correspondence between ground states of
$H^\perp$ and of $H=\sum h_i$; in particular, the constructed RVB parent
Hamiltonian has a topology-dependent degeneracy of the ground space
(four-fold on the torus), which is obtained from the ground space of the
dimer Hamiltonian by applying $\mc O_{\ds}^{\otimes K}$.

The Hamiltonian constructed this way is rather large: In particular, those
triangle terms $h_\medtriangleup^\perp$ which lie at the intersection of
three stars give rise to terms acting on three stars in
Fig.~\ref{fig:star-covering}.  In order to obtain simpler terms, one can
choose to adapt the framework for parent Hamiltonians of $G$--injective
PEPS derived in Ref.~\onlinecite{schuch:peps-sym} to the RVB state, which
results in a parent Hamiltonian acting on two overlapping stars, i.e.,
$19$ spins; we give the full derivation in
Appendix~\ref{appendix:twostars}.

\subsection{Hamiltonian for the dimer-RVB interpolation
\label{sec:ham-dimer-rvb-interpol}}

The relation between the dimer and RVB Hamiltonian extends to the whole
interpolating path introduced in Sec.~\ref{sec:dimer-rvb-interpol}, and
gives rise to a continuous path of local Hamiltonians interpolating between
the dimer model and the RVB state: In order to ensure continuity, we first
replace the dimer Hamiltonian 
$H^\perp=\sum_i h_i^\perp$
by an equivalent Hamiltonian
$\hat H^\perp :=
\sum_i \hat h_i^\perp + \sum_j \Pi_{j}$.
Here, $\hat h_i^\perp$ is the projection of $h_i^\perp$ onto 
$(\rg\,\mc P^\perp_{\ds})^{\kappa(i)}$, where the tensor product
$\otimes \kappa(i)$ goes over all stars supporting $h_i$ and we omit
vanishing $\hat h_i^\perp$, and $\Pi_j$ is the projector onto
$\rg\,\mc P_{\ds}^\perp$ on star $j$, where the sum ranges over all stars. 
As $H^\perp$ is a parent Hamiltonian, the restriction to $\rg\, \mc
P_{\ds}^\perp$ is ensured by the terms in $H^\perp$ locally; thus, $\hat
H^\perp$ has the same ground state space as $H^\perp$ with a spectal gap
above, and we can smoothly interpolate between $H^\perp$ and $\hat
H^\perp$ without changing the ground state space or closing the gap.  For
$\hat H^\perp$, it is now straightforward to construct a continuous
interpolating path $\hat H(\theta) = \sum_i \hat h_i(\theta) +
\Pi_j(\theta)$, where $\hat h_i(\theta) = \mc
O^{\otimes\kappa}_{\ds}(\theta) \hat h_i \mc O^{\otimes
\kappa}_{\ds}(\theta)$, and $\Pi_j(\theta)$ projects onto
$\rg\,P_{\ds}(\theta)$ for star $j$.

\section{Conclusions}

In this paper, we have applied the PEPS formalism to the study of the
Resonating Valence Bond states and dimer models.  We have discussed PEPS
representations of the RVB and the dimer model and studied their structure
and relation.  In particular, we have given a local unitary mapping
between the dimer model and the toric code;  furthermore, by defining the
dimer state with locally orthogonal dimers, we were able to devise a local
reversible mapping between the dimer state and the RVB state which allowed
us to prove that the RVB state is the four-fold degenerate ground state of
a local parent Hamiltonian for any finite lattice.  Subsequently, this
allowed us to devise a smooth interpolation between the dimer state and
the RVB state and the corresponding Hamiltonians.  We have studied this
interpolating path numerically, considering the behavior of correlation
functions, the rate at which the ground state changes, and the
entanglement spectrum, and have found that all of these quantities behave
smoothly and show no sign of a phase transition.

Our results make heavy use of the formalism of PEPS and their associated
parent Hamiltonians, and in particular of $G$--injectivity and
$G$--isometry, which we generalize in order to assess the RVB state on the
kagome lattice.  Similarly, the PEPS representation of the dimer--RVB
interpolation allowed for efficient numerical simulations, enabling us to
study the phase of the RVB state. We believe that these techniques will be
of further use in the study of related systems.  In particular, 
the results can be generalized to other lattices: Firstly, the PEPS
description of RVB states applies to arbitrary graphs.  All
lattices with the ``linear independence property''~\cite{chayes:linindep,%
seidel:kagome,wildeboer:linear-independence} are $G$--injective, and
whenever the linearly independent blocks allow cover the lattice up to
disconnected patches (such as in Fig.~\ref{fig:star-covering}), this
allows to interpolate between the RVB and the corresponding dimer state.
(This is the case, for instance, for the square-octagon or the star lattice
in Ref.~\cite{wildeboer:linear-independence}, whereas for the hexagonal
lattice it appears that no such covering can be found.)  Further, if the
dimer model can be expressed in terms of $\mathbb Z_2$--injective tensors,
this allows to conclude that it is equivalent to the toric code. (This is
the case e.g.\ for the star lattice, but not for the square-octagon
lattice.) Also note that our findings are not restricted to
spin-$\tfrac12$ $\mathrm{SU}(2)$ singlets as dimers, but can be
generalized to higher dimensional singlets or any other state, as long as
$\mathbb Z_2$--injectivity can be established.

\subsection*{Acknowledgements}
We acknowledge helpful comments by S.\ Kivelson and A.\ Seidel.  We wish
to thank the Perimeter Institute for Theoretical Physics in
Waterloo, Canada, and the Centro de Ciencias Pedro Pascual in Benasque,
Spain, where parts of this work were carried out, for their hospitality.
NS acknowledges support by the Alexander von Humboldt foundation, the
Caltech Institute for Quantum Information and Matter (an NSF Physics
Frontiers Center with support of the Gordon and Betty Moore Foundation)
and the NSF Grant No.~PHY-0803371.  
DP acknowledges support by the ``Agence Nationale de la Recherche" under
grant No.~ANR~2010~BLANC 0406-0, and CALMIP (Toulouse) for super-computer
ressources under Project P1231.
JIC acknowledges the EU project QUEVADIS, the DFG Forschergruppe 635, and
Caixa Manresa.  
DP-G acknowledges QUEVADIS and Spanish grants QUITEMAD and MTM2011-26912.  

\onecolumngrid

\twocolumngrid

\appendix

\quad 

\clearpage

\section{Different PEPS representations for the RVB and dimer state
\label{appendix:rvb-peps-representations}}

In this appendix, we give an overview of different PEPS representations of
the RVB and dimer state and how they are related.  We will start from the
representation introduced in Ref.~\onlinecite{verstraete:comp-power-of-peps}, 
and subsequently show how to derive from it the representation used in
this paper.

\subsection{PEPS representation of the RVB
\label{sec:appendix-pepsreps-standardrep}
}

\begin{figure}[b]
\begin{center}
\includegraphics[width=0.6\columnwidth]{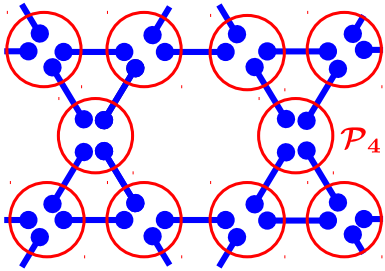}
\end{center}
\vspace*{-0.2cm}
\caption{\label{appb-fig:rvb-4bonds-peps}
PEPS representation for the RVB.  Place maximally entangled ``bonds''
$\ket{01}-\ket{10}+\ket{22}$ along the edges of the lattice (blue), and
subsequently apply the linear map (``PEPS projector'') $\mc P_4$ (red
circle), Eq.~(\ref{appb-eq:rvb-peps-proj}), which maps the four qutrits at each
vertex (encircled) to a spin-$\tfrac12$ degree of freedom. Alternatively,
we can obtain the (orthogonal) dimer state by replacing $\mc P_4$ with
$\mc P_{4,\perp}$, Eq.~(\ref{appb-eq:p-orth}): $\mc P_{4,\perp}$ ensures
that only one qutrit per vertex holds a singlet, but keeps the full
Hilbert space, thus ensuring local orthogonality of different dimer
configurations.
}
\end{figure}

We first explain the PEPS representation of the RVB state introduced in
Ref.~\onlinecite{verstraete:comp-power-of-peps}, which is illustrated in
Fig.~\ref{appb-fig:rvb-4bonds-peps}: We place states
\begin{equation}
        \label{appb-eq:rvb-bond}
\ket\omega = \ket{01}-\ket{10}+\ket{22}
\in\mb C^3\otimes \mb C^3
\end{equation}
along all edges of the lattice (observing its orientation);
this associates four three-level
systems (qutrits) with each vertex.  Then, we apply the following map to
the qutrits at each vertex, which maps them to one physical qubit:
\begin{equation}
\label{appb-eq:rvb-peps-proj}
\begin{aligned}
\mc P_4 =& \ket{0}\big(\bra{0222}+\bra{2022}+\bra{2202}+\bra{2220}\big)\\
     &+\ket{1}\big(\bra{1222}+\bra{2122}+\bra{2212}+\bra{2221}\big)\\
    =& \sum_{k=1}^4\;(\ket{0}\bra{0}_k+\ket{1}\bra{1}_k)\otimes
    \bra{222}_{/k}\ .\\
\end{aligned}
\end{equation}
Here, in the second formulation the sum runs over the four virtual systems
$k$, $\bra{0}_k$ and $\bra{1}_k$ act on the virtual system $k$, and
$\bra{222}_{/k}$ acts on all virtual systems but $k$.

What is the intuition underlying this construction?  The Hilbert
space holding the bond state, $\ket{\omega}=\ket{01}-\ket{10}+\ket{22}$,
Eq.~(\ref{appb-eq:rvb-bond}), should be understood as a direct sum
$\mathbb C^2\oplus \mathbb C^1$ [with a corresponding $\mathrm{SU}(2)$
representation $\tfrac12\oplus 0$].  The two-dimensional subspace, $\mf
S=\mathrm{span}\{\ket0,\ket1\}$, provides the spin-$\tfrac12$ degree of
freedom which holds the singlets $\ket{01}-\ket{10}$ used to construct the
RVB state, while the third level, $\ket{22}$, is used as a tag indicating
that there is no singlet along that edge, synchronizing this choice for
both ends.  In order to understand what happens when we apply $\mc P_4$ to
the bonds $\ket\omega$,  let us rewrite $\mc P_4$ as 
\[
\mc P_4 = \mc P_4\; \mc P_{4,\perp}\ ,
\]
where $\mc P_{4,\perp}$ is a projector acting \emph{within} the virtual
space as 
\begin{equation}
\label{appb-eq:p-orth}
\mc P_{4,\perp} = 
 \sum_{k=1}^4 \openone_{\mf S,k}\otimes
    \ket{222}_{/k}\bra{222}_{/k}\ ,
\end{equation}
with $\openone_{\mf S,k}$ the projector onto the subspace $\mf S$
on site $k$.
Thus, $\mc P_{4,\perp}$ projects \emph{one} virtual qutrit onto the ``singlet
space'' $\mathfrak S$, while the others are projected onto the $\ket2$
subspace.  Due to the form of $\ket\omega$, it follows that both sides of
each bond have to be projected onto the same subspace, i.e., either the
singlet is selected, or both ends of the bond need to be in the $\ket{2}$
state.  From these two observations, it follows that by applying $\mc
P_{4,\perp}$ to the bonds $\ket\omega$, we obtain configurations with singlets
along the edges, such that exactly one singlet is incident to each vertex,
while all other edges are in the ``unused'' state $\ket{22}$.  This way,
all possible dimer configurations are orthogonal---we can infer the
location of the dimers by measuring whether we are in the $\mf S$ subspace
or in the $\ket2$ state. Moreover, the final state has an equal weight of
all dimer configurations, as $\ket{01}-\ket{10}$ and $\ket{22}$ appear
with the same relative weight everywhere. Thus, it follows that the state
obtained by applying the maps $\mc P_{4,\perp}$ to the bonds $\ket\omega$
is a realization of the (orthogonal) dimer state.

It is now straightforward to see that applying $\mc P_4$ to this dimer
state gives the actual (non-orthogonal) RVB state: $\mc P_4$ keeps the
singlet degree of freedom at every vertex, but it (coherently) erases the
information about the edge the singlet is associated to; as the phases
between different dimer locations are chosen to be $+1$, we indeed obtain
an equal weight superposition of all dimer coverings with singlets, i.e.,
the RVB state.

\subsection{Simplified PEPS representation
\label{sec:simplified-rvb-PEPS}}

\begin{figure}[t]
\includegraphics[width=0.95\columnwidth]{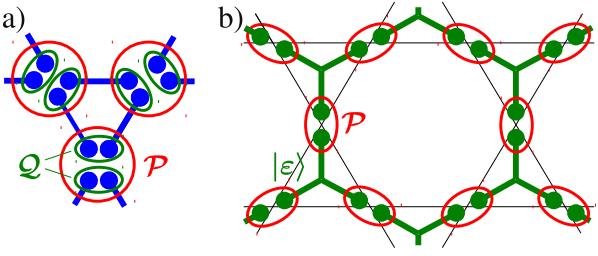}
\caption{\label{appb-fig:rvb-eps-p2-peps}
Simplified PEPS construction for the RVB and dimer state.  \textbf{a)}
Start from the representation Fig.~\ref{appb-fig:rvb-4bonds-peps} and
rewrite $\mc P_4$ in two layers, $\mc P_4=\mc P(\mc Q\otimes \mc Q)$,
Eq.~(\ref{appb-eq:P-P2Q-dec}). The $\mc Q$ act as indicated by the green
circles, and make sure \emph{at most} one singlet degree of freedom is
kept; they thus return one qutrit each.  $\mc P_2$ makes sure
\emph{exactly one} singlet is kept, which can come from either of the two
$\mc Q$'s.  \textbf{b)} The simplified PEPS construction: Applying three
$\mc Q$'s to three bonds around a loop results in a state $\ket\epsilon$
which holds at most one singlet, Eq.~(\ref{appb-eq:eps-bond}), and which is
placed across every triangle. Subsequently, $\mc P$ is applied to 
pick one of the singlets at each vertex.} \end{figure}

We will now prove that the PEPS representation of the RVB state
Eqs.~(\ref{appb-eq:rvb-bond},\ref{appb-eq:rvb-peps-proj}) is equivalent to
the form introduced in Sec.~\ref{sec:RVB-peps-representation}.  We start
by rewriting $\mc P_4$, Eq.~(\ref{appb-eq:rvb-bond}), as 
\begin{equation}
\label{appb-eq:P-P2Q-dec}
\mc P_4 = \mc P\;(\mc Q\otimes \mc Q)\ .
\end{equation}
Here, $\mc P:\mb C^3\otimes \mb C^3\rightarrow \mb C^2$ is defined just
as $\mc P_4$, but acting only on two virtual systems,
\[
\mc P = \ket{0}(\bra{02}+\bra{20})+\ket{1}(\bra{12}+\bra{21})
\]
[cf.~Eq.~(\ref{eq:rvb-proj})],
and $\mc Q:\mb C^3\otimes \mb C^3\rightarrow \mb C^3$ is defined as
\[
\mc Q = \ket{0}\big(\bra{02}+\bra{20}\big)+
    \ket{1}\big(\bra{12}+\bra{21}\big) + 
    \ket{2}\bra{22}\ ;
\]
it is straightforward to check that (\ref{appb-eq:P-P2Q-dec}) holds. We choose
to apply the $\mc Q$'s as depicted in Fig.~\ref{appb-fig:rvb-eps-p2-peps}a:
Then, applying $\mc Q^{\otimes 3}$ to $\ket{\omega}^{\otimes
3}$ across each
triangle yields a $3$-qutrit state
\begin{equation}
        \label{appb-eq:eps-bond}
\ket{\varepsilon} = \sum_{i,j,k=0}^3 \varepsilon_{ijk}\ket{ijk} + \ket{222}\ ,
\end{equation}
where $\varepsilon_{ijk}$ is the completely antisymmetric tensor with
$\varepsilon_{012}=1$, where $i$, $j$, and $k$ are oriented clockwise.
Thus, we obtain a PEPS construction for the RVB state where we place
tripartite bond states $\ket\varepsilon$ across triangles, and apply $\mc
P$ to each adjacent pair of qutrits, as shown in
Fig.~\ref{appb-fig:rvb-eps-p2-peps}b.

As already explained in Sec.~\ref{sec:dimer-peps-rep}, this PEPS
representation of the RVB state can again be understood as arising from a
PEPS representation of the dimer state, $\mc P = \mc P \; \mc P_{\perp}$
with 
        $\mc P_{\perp} = 
 \openone_{\mf S}\otimes \ketbra{2}
 + \ketbra{2}\otimes \openone_{\mf S}$.
Note that different dimer coverings are again locally orthogonal, although
in a different fashion---they become orthogonal on triangles.  Note that
this dimer state can be obtained from the one discussed in the previous
section~\ref{sec:appendix-pepsreps-standardrep} using that $\mc
P_{\perp}(\mc Q\otimes \mc Q)= (\mc R\otimes \mc R)\mc P_{4,\perp}$, where
$\mc R = \openone_{\mf S}\otimes \bra{2} + \bra{2} \otimes \openone_{\mf
S} + \ket{2}\bra{22}$: While $\mc R$ removes some degrees of freedom,
different dimer configurations remain orthogonal on triangles.

\section{\label{appendix:overlap-change}
Change of overlap and correlation functions
}

In this appendix, we show how to compute the change of the overlap of the
state along the dimer--RVB interpolation from two-point correlation
functions; it can be seen as a Hamiltonian-free PEPS version of the result
that the change of the overlap can be expressed as an integral over
imaginary time correlation
functions.~\cite{campos-venuti:overlap-correlationfunction} While we will
illustrate the derivation for the particular case of the dimer--RVB
interpolation, it will equally apply to most other PEPS interpolations.

The interpolating path is given by
\begin{equation}
        \label{appovl:eq1}
\ket{\psi_\theta} = \mc P_\theta^{\otimes N}
\ket{\Omega}\ ,
\end{equation}
with $\mc P_\theta$ as in Eq.~(\ref{eq:P-interpolation}), and where
$\ket\Omega\equiv\ket{\Psi_{\mathrm{dimer}}}$ is the dimer state.  
$\mc P_\theta$ can be split as
\begin{equation}
        \label{appovl:eq2}
\mc P_\theta = P_+ + \theta P_-\ ,
\end{equation}
where $P_+$ and $P_-$ are projectors independent of $\theta$,
$P_++P_-=\openone$; the following derivation holds for any interpolation
of the form Eqs.~(\ref{appovl:eq1},\ref{appovl:eq2}). Note that from
(\ref{appovl:eq2}), we have that
\begin{equation}
        \label{appovl:eq3}
\mc P_{\theta+d\theta} = (\openone+\tfrac1\theta P_- d\theta) \mc P_\theta\ .
\end{equation}
We want to compute the change of the normalized overlap 
\[
O(\theta,\theta+d\theta)=
\frac{\langle\psi_\theta\ket{\psi_{\theta+d\theta}}}{
\sqrt{\langle\psi_\theta\ket{\psi_{\theta}}}
\sqrt{\langle\psi_{\theta+d\theta}\ket{\psi_{\theta+d\theta}}}}\ .
\]
As the first-order change to $O(\theta,d\theta)$ can be at most a
(unphysical) phase change, we are interested in the second order in
$d\theta$.  We write $\ket{\psi_{\theta+d\theta}} =
\ket{\psi_\theta}+\ket{d\psi_\theta}$, and from 
Eqs.~(\ref{appovl:eq1}) and (\ref{appovl:eq3}), it follows that
$\ket{d\psi_\theta}$ can be expanded in orders of $d\theta$.

For notational convenience, let us define
\begin{align*}
\mc N &= \langle\psi_\theta\ket{\psi_{\theta}}\\
A &= \langle \psi_{\theta}\ket{d\psi_{\theta}}/\mc N\\
B &= \langle d\psi_{\theta}\ket{d\psi_{\theta}}/\mc N\ .
\end{align*}
Note that since both $\ket\Omega$ and $\mc P_\theta$ in
(\ref{appovl:eq1}) are real, $\ket{\psi_\theta}$ is real, and thus $A$ is
real;  we will use this in
the following derivation.  Using the above definitions (where $A$ is first
order and $B$ second order in $d\theta$) we have (neglecting third and
higher order terms)
\begin{align*}
O(\theta,\theta+d\theta) &= 
\frac{\mc N(1+A)}{\mc N \sqrt{1+(2A+B)}} 
\\
&\approx 
(1+A)\left[1-\frac{2A+B}{2}+\frac{3(2A)^2}{8}\right]
\\
&=
1-\tfrac12(B-A^2)\ .
\end{align*}

Clearly, 
\[
\ket{d\psi_\theta} = 
\sum \frac{\mathrm{d} \mc P_\theta}{\mathrm{d}\theta} 
    \mc P^{\otimes(N-1)}_\theta \ket\Omega d\theta
= \sum_i \frac{1}{\theta} P^i_- \ket{\psi_\theta}\,d\theta\ ,
\]
where the sum is over all possible positions $i$ of the $\tfrac1\theta
P_-^i$.
(As $A$ only appears as $A^2$ in the overlap, we can neglect the second
order term.)
It follows  that 
\[
A=
    \frac1\theta\sum_i \frac{\bra{\psi_\theta}P_-^i\ket{\psi_\theta}}{\mc
            N}\,d\theta
\equiv 
    \frac1\theta\sum_i \langle P_-^i\rangle_\theta\,d\theta\ ,
\]
where we have used the notation 
\[
\langle X \rangle_\theta := 
\frac{\bra{\psi_\theta} X \ket{\psi_\theta}}{
\langle \psi_\theta\ket{\psi_\theta}}\ .
\]
On the other hand, to second order
\[
B=\frac{1}{\theta^2} 
\sum_{ij} \langle P_-^i P_-^j\rangle_\theta
\,(d\theta)^2\ ,
\]
and thus, the second-order term in the expansion 
\[
O(\theta,\theta+d\theta) = 1-\tfrac12 g_\theta N (d\theta)^2
\]
can be expressed as an average over connected correlation functions:
\begin{equation}
        \label{appovl:eq4}
g_\theta 
    = \frac{B-A^2}{N(d\theta)^2} 
    =\frac{1}{\theta^2N} 
    \sum_{ij} \langle \hat P_-^i \hat P_-^j\rangle_\theta 
\end{equation}
where 
$\hat P_-^i := P_-^i - \langle P_-^i\rangle_\theta$, and $N$ is the 
number of sites.  
This derivation shows that for PEPS interpolations, we can infer the
change of the overlap from two-point correlators; this argument
can be extended to relate the $p$-th order expansion coefficient of the
overlap to $p$-point correlators.

For systems with finite correlation length $\xi$, the sum in
Eq.~(\ref{appovl:eq4}) is proportional to $\xi^2N$, and thus,
$g_\theta\sim \xi^2/\theta^2$. Therefore, the only way $g_\theta$ can
diverge when the correlation functions decay exponentially is if 
$\theta\rightarrow0$. Note, however, that for $\theta=0$, $\langle \hat
P_-^i\hat P_-^j\rangle_\theta=0$, so whether $g_\theta$ diverges at
$\theta\rightarrow0$ depends on the rate at which the latter vanishes.

Note that Eq.~(\ref{appovl:eq4}) does not apply when $\theta=0$. For that
case, one can easily check that
$\bra{\psi_{\theta=0}}d\psi_{\theta=0}\rangle = 0$ and thus $A=0$. On the
other hand, $\bra{d\psi_{\theta=0}}d\psi_{\theta=0}\rangle/(d\theta)^2$ is
the norm of the PEPS where $\mc P_{\theta=0}\equiv P_+$ has been replaced
by $P_-$ at a single site (summed over all sites), which immediately shows
how to compute $g_0=B/N(d\theta)^2$.

\section{\label{appendix:oRVB-parent}
A simple parent Hamiltonian for the orthogonal RVB}

In this appendix, we show how to directly derive a parent Hamiltonian for
the dimer state (in the representation of Sec.~\ref{sec:dimer-peps-rep})
by relating it to the toric code
on the same lattice. 

\begin{figure}
\includegraphics[width=\columnwidth]{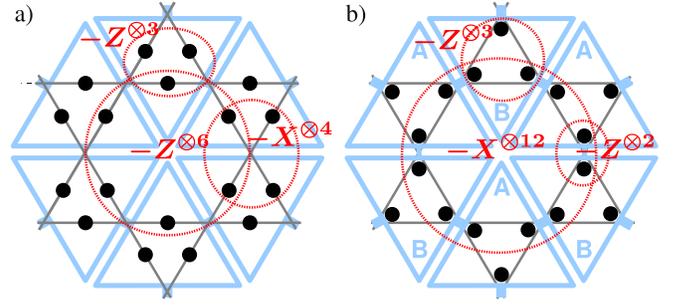}
\caption{\label{fig:tc-kagome-ham}
 \textbf{a)}
Tensor network and Hamiltonian for the toric code on the kagome lattice.
Two types of terms ensure the parity constraints over the
triangular and hexagonal plaquettes, while the vertex term enforces an equal
weight superposition of all even parity configurations. Note that the
$-Z^{\otimes 3}$ acting on the triangles can be understood as enforcing
each individual tensor to be in the correct subspace.
\textbf{b)}
Hamiltonian terms obtained by transforming each tensor unitarily into a
projector onto the symmetric subspace. $Z^{\otimes 3}$ still enforces that
the tensor is in the correct subspace; further, there is a $Z^{\otimes 2}$
at each vertex enforcing equality of the adjacent qubits---translated to
the dimer state, this will enforce that there is exactly one singlet
adjacent to each vertex.  Finally, the $X^{\otimes 12}$ term around the
hexagon makes sure that all such configurations appear with equal
weight---it will make the dimer configurations resonate in the orthogonal
RVB state. (Note that this model can also be understood as a toric code
model on a ``decorated'' hexagonal lattice with two sites per edge and a
vertex between, and $X$ stabilizers around plaquettes and $Z$ stabilizers
across vertices.)
}
\end{figure}

Let us start by considering the toric code model on a kagome lattice,
where the spins sit on the edges of the lattice, as illustrated in
Fig.~\ref{fig:tc-kagome-ham}a.  The toric code Hamiltonian has two types
of terms:  First, it has terms acting on the four qubits across each
vertex, as
\[
h^{TC}_v = -X^{\otimes 4}\ ,
\]
and second, it has terms acting across plaquettes---triangles and
hexagons---as 
\[
h^{TC}_\medtriangleup =
h^{TC}_\medtriangledown = -Z^{\otimes 3}\ ,\quad 
h^{TC}_{\hexagon} = -Z^{\otimes 6}\ ,
\]
where $X$ and $Z$ are the Pauli matrices. For future purposes, we have
introduced different labels for the two types of triangles:
$h^{TC}_\medtriangleup$ refers to an upward pointing \emph{tensor}
(sketched blue in Fig.~\ref{fig:tc-kagome-ham}a), and thus to a downwards
triangle, and vice versa.
In analogy to the square lattice toric code, Eq.~(\ref{eq:tc-proj}), we
can write the toric code state with triangular $\mathbb Z_2$--invariant
tensors
\[
\mc P^{\mathrm{TC}} = \sum_{\alpha,\beta,\gamma} = 
\ket{\alpha-\beta,\beta-\gamma,\gamma-\alpha}\bra{\hat\alpha,\hat\beta,\hat\gamma}\
.
\]
The states $\bra{\hat\eta}$ on the virtual level are again in the $x$
eigenbasis, which makes them invariant under $Z^{\otimes 3}$. 
The tensors are marked blue in Fig.~\ref{fig:tc-kagome-ham}a, with the
virtual indices associated to vertices of the kagome lattice.
$\mc P^{\mathrm{TC}}$ can be transformed by a unitary acting on the
physical system to the projector onto the invariant subspace,
\[
\tilde{\mc P}^{\mathrm{TC}} = 
    \tfrac12(\openone^{\otimes 3} + Z^{\otimes 3})\ ,
\]
where the physical qubits are now associated to vertices,
cf.~Fig.~\ref{fig:tc-kagome-ham}b.  It
can be easily checked  that the
action of the Hamiltonian terms in the transformed basis is as follows
(illustrated in Fig.~\ref{fig:tc-kagome-ham}b):
\[
\tilde h^{\mathrm{TC}}_v = - Z^{\otimes 2}
\]
enforces that the two qubits per vertex are in the same state (namely
equal to the value of the virtual index);
\[
\tilde h^{\mathrm{TC}}_{\hexagon} = -X^{\otimes 12}
\]
acts on the $12$ qubits around each hexagon and ensures that all bond
configuration appear with the same probability; finally,
in the original representation,
$h^{\mathrm{TC}}_\medtriangleup=h^{\mathrm{TC}}_\medtriangleup$ enforced
that the qubits at each tensor
are in the proper subspace, it thus has to be replaced by the
corresponding term for the transformed tensor,
\[
\tilde h^{\mathrm{TC}}_\medtriangleup = 
\tilde h^{\mathrm{TC}}_\medtriangledown = -Z^{\otimes 3}\ .
\]
The three terms are depicted in Fig.~\ref{fig:tc-kagome-ham}b.

Let us now consider what happens if we change all $A$--type tensors
(upwards pointing triangles) to projectors onto the antisymmetric
subspace,
\[
\tilde{\mc P}^{\mathrm{def}}_\medtriangleup = 
\tfrac12(\openone^{\otimes 3}-Z^{\otimes 3})\ ,
\]
while the $B$--type tensors (downwards pointing triangles) remain of the
form
\[
\tilde{\mc P}^{\mathrm{def}}_\medtriangledown = 
\tfrac12(\openone^{\otimes 3}+Z^{\otimes 3})\ .
\]
Clearly, this gives rise to a different triangular Hamiltonian for all 
$A$--type tensors,
\[
\tilde h^{\mathrm{def}}_\medtriangleup = Z^{\otimes 3}\ ,
\]
which enforces odd parity, while we keep $ \tilde
h^{\mathrm{def}}_\medtriangledown = \tilde
h^{\mathrm{TC}}_\medtriangledown$ for the $B$--type tensors.  All other
Hamiltonian terms remain unchanged, since we still need to ensure that the
qubits on both sides of the vertex are equal 
($\tilde h^{\mathrm{def}}_v=\tilde h^{\mathrm{TC}}_v$),
and that all configurations appear in an equal weight superposition
($\tilde h^{\mathrm{def}}_{\hexagon}=\tilde h^{\mathrm{TC}}_{\hexagon}$).

\begin{figure}[t]
        \includegraphics[width=\columnwidth]{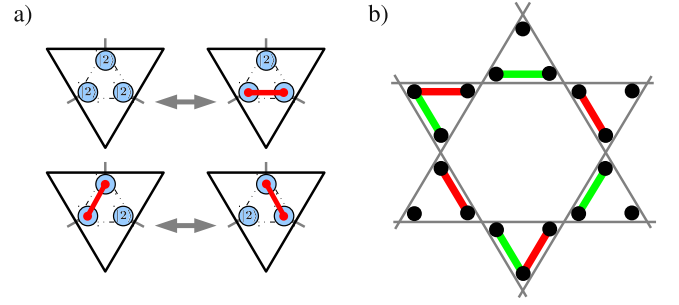}
\caption{\label{fig:xi-resonance}
Resonance moves in the dimer state. \textbf{a)} Resonance move
implemented by a single $\Xi$, Eq.~(\ref{eq:xi-resonance-def}): $\Xi$
acts on the lower two qutrits only and flips the two configurations
singlet/no singlet, and the two configurations singlet left/singlet right.
\textbf{a)} Applying $\Xi^{\otimes 6}$ around the hexagon gives a
resonance between pairs of dimer configurations around the hexagon (marked
red and green, respectively).  Note that any such pair of dimer
configuration can be described by a closed loop of even length on the
star, cf.~Ref.~\onlinecite{seidel:kagome}.}
\end{figure}

The projectors $\tilde{\mc P}^\mathrm{def}_\medtriangleup$ and $\tilde{\mc
P}^\mathrm{def}_\medtriangledown$ can now be transformed to the $A$ and
$B$ tensors for the dimer state by local isometries, which allows us to
explicitly construct its parent Hamiltonian: First, the terms $\tilde
h^{\mathrm{def}}_\medtriangleup$ and $\tilde
h^{\mathrm{def}}_\medtriangledown$ enfore that the sites at each tensor
are in the correct subspace; they are thus mapped to a local term
$h_\medtriangledown = h_\medtriangleup$ which have the four states of
Fig.~\ref{fig:triangle-configs} as ground states. The vertex term 
$\tilde h^{\mathrm{def}}_v$ is transformed to 
\[
h_v = -(\openone_{\mf S}\otimes \ketbra{2}
    +\ketbra{2}\otimes\openone_{\mf S})\ ,
\]
$\mathfrak S=\mathrm{span}\,\{\ket0,\ket1\}$,
which ensures there is only one singlet adjacent to each vertex. Finally,
$\tilde h^{\mathrm{def}}_{\hexagon}$ is transformed to a term which swaps the
singlet/no-singlet configuration for the $12$ qutrits adjacent to each
hexagon, observing the orientation of the singlets: This is achieved by an
operator
\[
h_{\hexagon} = -\Xi^{\otimes 6}\ ,
\]
where $\Xi$ acts on the two qutrits in each triangle adjacent to the
hexagon as
\begin{equation}
        \label{eq:xi-resonance-def}
\Xi = \ket{22}(\bra{01}-\bra{10}) - \ket{20}\bra{02} - \ket{21}\bra{12} +
\mathrm{h.c.}\ ,
\end{equation}
with proper orientation of the singlet---the first term changes between a
singlet and no singlet along the edge, and the latter two terms flip the
location of a (outwards-pointing) singlet in the triangle; the negative
sign is due to the fact that the two singlets are oriented differently.
The action of $\Xi$ is illustrated in Fig.~\ref{fig:xi-resonance}a.  Note
that this $\Xi^{\otimes 6}$, although it acts on a hexagon only, in fact
achieves resonance between all loop configurations on the star built
around the hexagon,
as illustrated in Fig.~\ref{fig:xi-resonance}b for a particular instance.

\section{\label{appendix:twostars}
$G$--injective toolbox and parent Hamiltonian for the RVB state}

In the main text, we have seen how to manipulate the RVB state in a kagome
lattice, to connect it with the dimer state and then with the toric code
in such a way that we can apply the results about $\Z_2$-injective PEPS
given in Ref.~\onlinecite{schuch:peps-sym} to give a local frustration
free Hamiltonian with topological features for the RVB state. Another
possible approach is to modify the toolbox for G-injective PEPS developed
in Ref.~\onlinecite{schuch:peps-sym} to
adapt it to  lattices different than the square one, even frustrated ones.
Since this route is necessarily technical, and assumes the reader is
familiar with Ref.~\onlinecite{schuch:peps-sym}, we only sketch it in
this appendix. The payoff: a more local Hamiltonian---acting on two
partially overlapping stars, that is, 19 spins---and the possibility of
applying the $G$--injectivity toolbox to rather general
lattices in the future.

For simplicity, we will stick to $\Z_2$-injectivity and the kagome
lattice, but it is straighforward how to extend the ideas to general
finite groups and general lattices. We will work with the tensor network
picture of the RVB state given in Fig.~\ref{fig:rvb-tn-description}b,
where the RVB state is recovered from contracting two types of tensors $P$
and $E$.  In the first part of this appendix, Sections \ref{subsec1} to
\ref{subsec3}, we will generalize the three key ingredients of Ref.
\onlinecite{schuch:peps-sym}: (i)~showing that $\Z_2$-injectivity is
stable while growing the lattice, (ii) showing the ``intersection
property'' for good-shaped Hamiltonians---as the one acting on three
overlapping stars---and (iii) showing that we recover the desired ground
space while closing the boundaries. All that will produce a local
frustration free Hamiltonian for the RVB state with local terms acting on
three overlapping stars and 4-fold degeneracy. Finally, in Section
\ref{subsec4}, we will see how to reduce the size of the local Hamiltonian
to two overlapping stars. For that, we will need the concept of linear
independence introduced by Seidel in Ref.~\onlinecite{seidel:kagome}.
\subsection{Growing $\Z_2$-injectivity}\label{subsec1}

\begin{figure}[t]
  \includegraphics[width=\columnwidth]{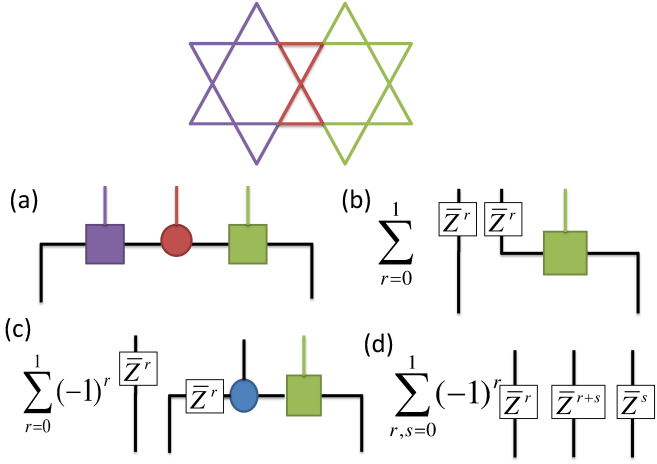}
\caption{Graphical proof that
$\Z_2$-injectivity is stable as growing a region. We illustrate it in the
case of growing from one to two partially overlapping stars. 
Our aim is to prove the existence of a left-inverse on the invariant
subspace on the two stars, using the existence of the left-inverse on a
single star.  Since we are considering the tensor network representation
of the RVB state made out of
tensors $P$---with physical index---and triangular links $E$ without
physical index, given a region, all one needs is to clarify whether the
outgoing auxiliary indices come from a $P$ tensor or from an $E$ tensor.
In our case, we start with two overlapping stars with outgoing indices of
type $P$ and divide them in three regions. The middle one (bow-tie) with
all outgoing indices of type $P$ and the rest. By gathering indices
together, we have figure (a), where the circle represents the bow-tie and
each square one of the other two regions. The crucial property is that the
circle together with {\it any} of the two squares is $\Z_2$-injective. We
proceed as follows. We apply the inverse to the circle and left square,
getting Figure (b), where $\bar{Z}$ now means $\otimes \bar{Z}$. The next
step, illustrated in (c) is to build a new circle tensor attached to the
right square tensor, and passing the $\bar{Z}^r$ through using the
symmetry of the circle tensor. Now we invert the remaining two tensors
getting (d). By tracing the middle (extra)  line we obtain $\delta_{rs}$
and therefore finish the argument since we have constructed the desired
inverse.}\label{fig2} \end{figure}

\begin{figure}[t]
  \includegraphics[width=\columnwidth]{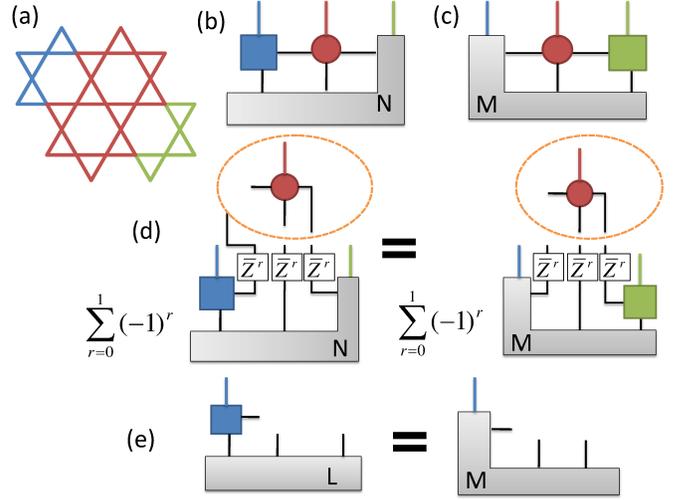}\\
  \caption{Graphical proof of the intersection property for the case
$h_1+h_2$ where $h_1$ acts on the 3 stars colored in blue-red in (a) and
$h_2$ acts on those colored in red-green. It is easy to see that, by
separating the common part (2 red stars) and the rest, with the convention
that the common part has all outgoing indices of type P, $\ker{h_1}\otimes
\Hc_\mathrm{green}$ is given by figure (b), where $N$ is an arbitrary
tensor, whereas $\Hc _\mathrm{blue}\otimes \ker{h_2}$ is given by figure
(c), where $M$ is arbitrary.  Here $\Hc_\mathrm{green}$ and
$\Hc_\mathrm{blue}$ represent the Hilbert spaces associated to the
physical indices in the green and blue regions, respectively. As before,
we have gathered together all virtual (as well as the physical) indices within
each type of tensor. Given a state $|\psi\>$ in the intersection of both
kernels, there exist $N$ and $M$ such that $\ket{\psi}$ can be expressed
both in the form (b) and (c).  In the equality (b)=(c) we invert
the circle tensor, obtaining the equality given in figure (d). After that,
we apply an auxiliary extra circle tensor, then the inverse of the
red-green tensors (3 stars), which is also $\Z_2$-injective by the
previous section, and then trace the extra virtual system. Since
one can assume wlog that both $M$ and $N$ have the symmetry $\otimes \bar{Z}$,
we can also eliminate the sums. After all that we arrive at figure (e) for
some tensor $L$. Substituting (e) in (c) gives the desired form for
$|\psi\>$ and hence the result.}\label{fig3} \end{figure}

We recall from the main text that a region in a tensor network is
$\Z_2$-injective if it is $\Z_2$-invariant, as illustrated in
Fig.~\ref{fig:z-invariant-tn},
 and in addition one can left-invert the tensor in that
region (understood as a map from the auxiliary to the physical system) to
get the projection onto the invariant subspace.  In
Figure~\ref{fig:pstar-definition} of the main
text, the basic $\Z_2$-injective block---one star---is depicted. In Figure
\ref{fig2} we illustrate how to grow the $\Z_2$-injective property from
one to two partially overlapping stars, but the procedure can be applied
to grow arbitrarily large regions. The only basic property we need for
that is the possibility of decomposing the final tensor as in Figure
\ref{fig2}a, and that this property is preserved as we grow the region.
The general result is that for regions that are grown  keeping this
property (it is trivial that one can grow in this way all regions that are
required to be $\Z_2$-injective later on in this appendix), there is an
inverse giving the projector onto the invariant subspace:
\begin{equation*} \frac{1}{2}\sum_{r=0}^1 (-1)^{|t|r} \bar{Z}^{r\,\otimes
|b|} ,\end{equation*} where $|t|$ is the number of tensors, both of $P$
and $E$ type, $|b|$ is the number of outgoing indices in the region, and
$$\bar{Z}:=\left( \begin{array}{ccc} 1 & &  \\ & 1 &  \\ &  & -1 \\
\end{array} \right) .$$ As commented in the main text, the extra sign one
obtains depending on the parity of the number of elementary tensors  in
the region only affects the definition of ``invariant subspace'', which
can be the symmetric ($|t|$ even) or the antisymmetric one ($|t|$ odd).

\subsection{Intersection property}\label{subsec2}

\begin{figure}[b]
\includegraphics[width=\columnwidth]{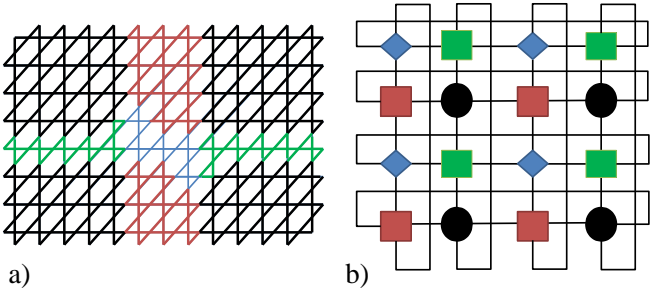}
\caption{The kagome lattice with periodic boundary conditions can be
decomposed as in (b), where all outgoing indices on the black regions are
of type $P$, whereas all outgoing indices of the blue regions are of type
$E$. All tensor in (b) have an additional physical index "going out of the
paper". A possible way to obtain such a decomposition is illustrated in
(a). The black regions are $\Z_2$-injective, the same happens if one takes
two black regions with the intermediate red or green one or if one takes
the four black regions with the intermediate red, green and blue ones. The
blue region is the one corresponding to the cross points. Note that the
black and the blue only interact through a red or green region, exactly as
required in (b).}\label{kagome} \end{figure}

We recall from the main text and Ref.~\onlinecite{schuch:peps-sym} that
given a PEPS, one can construct parent Hamiltonians $H=\sum_R h_R$ where
the local term $h_R$
only acts on a region $R$. The way to construct them is simple. Each $h_R$
is the orthogonal projector (or any positive semidefinite operator) whose
kernel coincides with the range of the PEPS tensor on that region (seen as
a map from the auxiliary to the physical indices in the region). This
Hamiltonian is $\ge 0$, frustration free and has the initial PEPS as
ground state. In Ref.~\onlinecite{schuch:peps-sym} we noticed how in the square lattice,
$G$-injectivity  preserves the structure of the parent Hamiltonian as we
grow it properly (i.e., by translational invariance) from a single local
operator. That is, the kernel of $\sum_{R\subset \mathcal{R}} h_R$ is {\it
exactly} the range of the PEPS tensor in the larger region $\mathcal{R}$.
This property is known as the ``intersection property'' since it is
related to the way the kernels of $h_R$ for different regions intersect.
We will illustrate here the intersection property for the case $h_1+h_2$
where $h_1$ acts on the 3 stars colored in blue-red in Figure
\ref{fig3}a and $h_2$ acts on those colored in red-green. The only
property one needs to make the argument work is that the common
interaction area is $\Z_2$-injective and that the non-common areas are not
directly connected. The proof is graphically given in Fig. \ref{fig3}.

\subsection{Closing the boundaries}\label{subsec3}

\begin{figure}[b]
\includegraphics[width=\columnwidth]{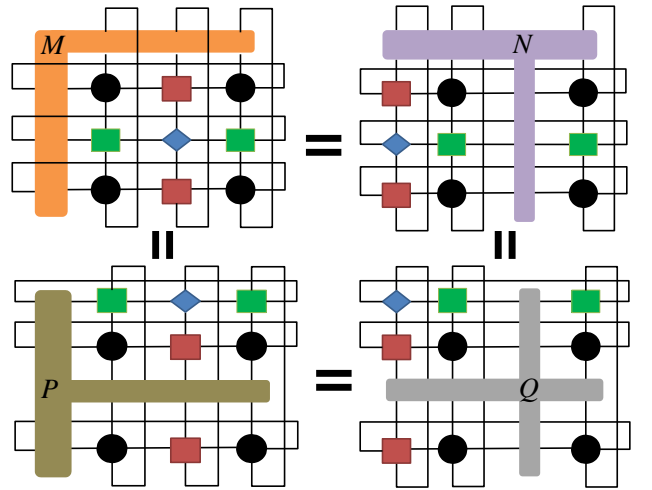}
\caption{Any state which
is in the ground state of the parent Hamiltonian when we close the
boundaries in a periodic way (where we close the boundaries as indicated
in Figure \ref{kagome}a) can be represented
graphically in any of the forms given in the figure. As in Figure
\ref{kagome}, each of the tensors, including the ``boundary conditions''
have a physical index going ``out of the paper''.}\label{clousure}
\end{figure}

It is not difficult to see that, if one considers the RVB state as a PEPS
in a torus of arbitrary size, by blocking tensors and gathering indices
together, one has the tensor network given in Figure \ref{kagome}b. It
may seem at first sight that we are back to a square-lattice PEPS.
However, there is a crucial difference. Not all tensors in the square
decomposition of Figure \ref{kagome}b are $\Z_2$-injective. We do
require that the four overlapping regions $\mathcal{R}_1,\ldots,
\mathcal{R}_4$ resulting from removing one row and one column with an
intersecting blue tensor contain all possible 3-star regions and that they
can be constructed by growing 3-star regions keeping the intersection
property commented on last section.  Figure \ref{kagome}a illustrates a
possible way of getting that in the surrounding of a blue tensor.
We will now use the trivial fact that, given the frustration free Hamiltonian 
\begin{equation}
\label{eq-parent}
H=\sum_{s\in  \{\rm{3-star}\}} h_s\end{equation} 
resulting from summing the local projectors corresponding to each 3-star
region in the torus, its ground space coincides with $\cap_{k=1}^4 \ker
{H_k}$, where $H_k=\sum_{s\subset \mathcal{R}_k} h_s$ is the Hamiltonian
corresponding to the region $\mathcal{R}_k$. By the intersection property
proved in last section, the associated kernels are given by Figure
\ref{clousure}.

\begin{figure}
  \includegraphics[width=7cm]{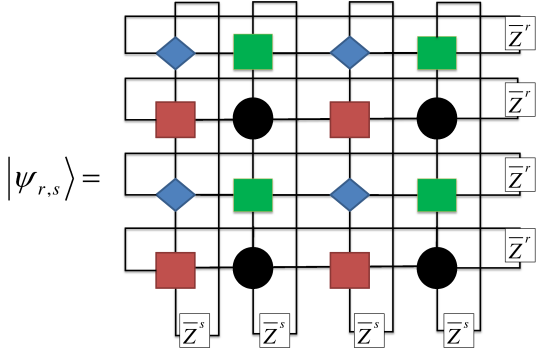}\\
  \caption{The ground space of the Hamiltonian with periodic boundary
conditions has dimension four and is spanned by the states $|\psi_{r,s}\>$
in the figure. As above, $\bar{Z}$ means $\otimes
\bar{Z}$.}\label{ground-state} \end{figure}

The argument now is a modification of Theorem 5.5 of
Ref.~\onlinecite{schuch:peps-sym}. We include a sketch for the sake of
completeness. The aim is to show that
the ground space of (\ref{eq-parent}) is spanned by the four states
$|\psi_{r,s}\>$, $r,s=0,1$, depicted in Figure \ref{ground-state}.

\begin{figure}[b]
  \includegraphics[width=\columnwidth]{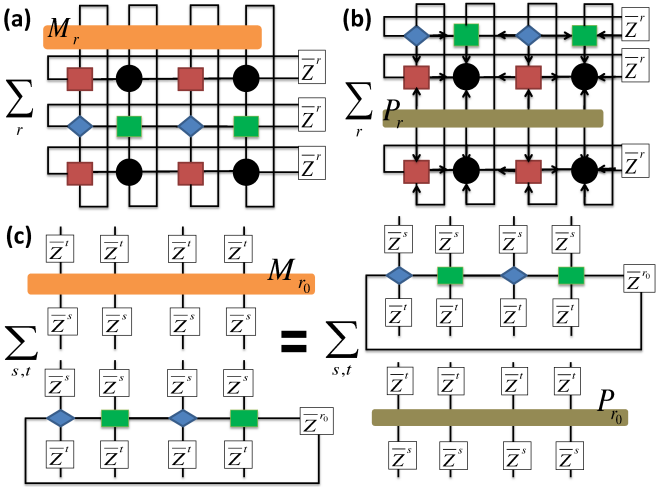}
  \caption{Intermediate step in the proof that the ground space of the
Hamiltonian is spanned by the vectors given in Figure \ref{ground-state}
above.}\label{fig4} \end{figure}

Using the two horizontal equalities of Figure \ref{clousure} and reasoning
as in the previous sections (inverting the two columns which are
$\Z_2$-injective, growing new tensors around the remaining column,
inverting them and taking traces), we obtain that the state is of the form
of Figure \ref{fig4}a (from the upper equality in Fig.~\ref{clousure}), as
well as of the form of Figure \ref{fig4}b (from  the lower equality).  The
vertical equalities then give that Fig.~\ref{fig4}a equals
Fig.~\ref{fig4}b.

We can now invert the two rows containing black tensors. The way to do it
is to invert the two black tensors with one intermediate red one, then
construct two new black tensors around the remaining red one, invert again
and take traces. This even allows us to select a particular $r=r_0$, 
this way removing the sum over $r$. We end up in Figure \ref{fig4}c and from
there, with similar techniques we show that the given state must be a
linear combination of states of the form given in Figure
\ref{ground-state}. Since clearly all such states belong to the ground
space of the Hamiltonian, and all of them are linearly independent, they
form a basis of the ground space. Note that, as in the toric code, all
four ground states are related by non-local string operators on the
virtual level. Indeed, the action of these virtual operators is nothing
but changing the sign (or flipping the direction) of the dimers crossing
the links in which they lie. Note also that the same procedure and
conclusions can be obtained starting from the parent Hamiltonian of a
basic region different from the 3-star one, as long as one can grow
properly the lattice starting from this region while keeping the
intersection property. Another example of such a basic region is the
7-star region of Figure \ref{2-star}a.

\subsection{A 2-star Hamiltonian}\label{subsec4}

\begin{figure}[t]
  \includegraphics[width=\columnwidth]{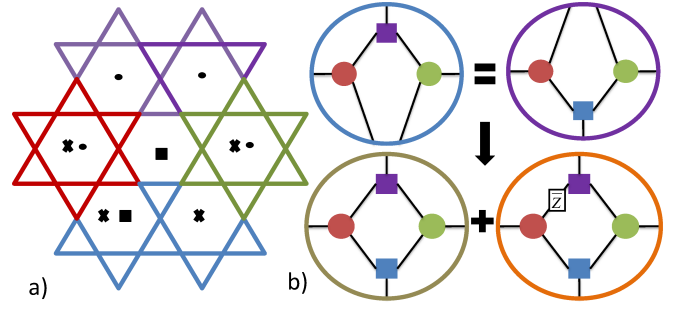}\\
  \caption{The figure illustrates the first step in the reduction to show
that the parent Hamiltonian on two overlapping stars suffices to give the
RVB state as a unique ground state -- up to the string operators given in
Figure \ref{ground-state} which induce a 4-fold degeneracy. The left
(right) upper part of (b) represents the kernel of the Hamiltonian
grown from 2-star regions which acts on the stars with a dot (cross). The
circles represent the stars with the same color in (a), with all outgoing
indices of type $P$. The squares represent the rest of (a),
having outgoing indices both of type $P$ (those connecting with the
boundary) and type $E$ (those connecting with the circles). As in the
previous figures, all bonds connecting the same objects have been gathered
together in the same bond. Also, all tensors have an additional physical
index which ``goes out of the paper'' The boundary conditions in the upper
part of (b) have also a physical index corresponding to the part of (a)
with the same color. The intersection of these two kernels, which is the
kernel of $H_6=\sum_{x,\circ} H_2$, is the sum of the two subspaces
depicted in the lower part of (b). Needless to say that the boundaries of
the lower part of (b) do not have physical index.}\label{2-star}
\end{figure}

The first step to show that a 2-star Hamiltonian suffices is sketched in
Figure \ref{2-star}. Starting from the parent Hamiltonian for 2-star
regions, and using the intersection property, one gets that the
Hamiltonian acting on the four stars marked with a dot (cross) has
a kernel equal to the upper-left (upper-right) subspace in
Fig.~\ref{2-star}b. Therefore, the kernel of the Hamiltonian
$H_6=\sum_{x,\circ}H_2$---the sum of all terms acting on two overlapping stars
marked by both circles or crosses---is the intersection of those
subspaces. Take a vector in the intersection. Then we can reason with the
very same techniques as above. First we invert the circles, then grow
again circles at both sides of the blue square and invert the joint tensor
circle-square-circle. Closing the unnecessary indices by taking traces we
arrive at the fact that the original tensor indeed belongs to the sum of
the subspaces in the lower part of Figure \ref{2-star}b. Since clearly
each one of these subspaces is contained in the kernel of $H_6$, the lower
part of Figure \ref{2-star} is indeed equal to the kernel of $H_6$.

The second step is to add a 2-star term including the middle star, for
instance, the one marked by squares in Figure \ref{2-star}a. We will
show that this term penalizes all the states in the lower-right subspace
of Figure \ref{2-star}b. That is, $\cap_{x,\circ,\square}\ker H_2=\ker
\{\sum_{x,\circ,\square} H_2\}$ equals the lower-left subspace in Figure
\ref{2-star}b, where the intersection/sum runs on all 2-star
Hamiltonians in the 7-star region given in Figure \ref{2-star}a. This
procedure illustrates how from 2-star Hamiltonians one reaches the desired
ground space in the 7-star region of Figure \ref{2-star}a. Since this
ground space is the same one obtains growing from 3-stars, we are done.

\begin{figure}[t]
\vspace{0.8cm}
\includegraphics[width=\columnwidth]{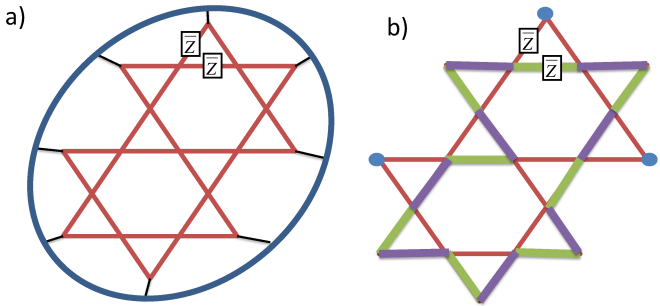} 
\caption{The figure illustrates the second step in the reduction to a
2-star Hamiltonian. (a) represents the space $\mathcal{S}$, coming from
the lower-right part of Figure \ref{2-star}b, each state of which we want
to penalize by the 2-star term marked by squares in Figure \ref{2-star}a.
All outgoing indices are of type $P$. (b) illustrates the idea of
``completing the path'': for each set of free sites --marked by a
circle--, there exist 2 disjoint dimer coverings (in green and violet
respectively) compatible with the choice of free sites, which in addition
form a Hamiltonian-cycle in the remaining graph.}\label{2-star-2}
\end{figure}

Let us first explain the intuition behind the following reasoning. For
this, we will use the PEPS representation of
the RVB of Fig.~\ref{appb-fig:rvb-4bonds-peps}.
The intuition is that the $\bar{Z}$ appearing in the
lower-right part of Figure \ref{2-star}b---which denotes nothing but a
$\bar{Z}^{\otimes 4}$ in the 4 bonds connecting the corresponding
regions---cannot escape the middle star: That is, no matter how we use the
gauge symmetry of the tensors in the PEPS, there will be one, and only
one, $\bar{Z}$ in one of the bonds of the middle hexagon. This
``excitation'' can be then detected by a 2-star term including the middle
star. To make it rigorous we will need the linear-independence property,
proven by Seidel in Ref.~\onlinecite{seidel:kagome}:
The sum $\oplus_D\{|\sigma(D)\>\otimes \mathcal{H}_{D,{\rm
free}}\}$ is direct. Here, $D$ means a fixed dimer covering of the
2-stars, $|\sigma(D)\>$  the associated tensor product of singlets,  and
$\mathcal{H}_{D,{\rm free}}$ the total Hilbert space of those {\it free}
sites which do not have any dimer. The sum runs on all possible dimer
coverings of the 2-stars.

Let us now take the Hamiltonian acting on the stars marked with a square
in Figure \ref{2-star}a. Using the $\Z_2$-injectivity, its kernel is
exactly given by $\mathcal{K}=\oplus_b \{\mathcal{H}_b\otimes\sum_{\rm
valid}|\sigma(D)\>\}$, where $b$ is a fixed choice of free sites---with
an odd number of terms to allow dimer coverings---and the sum runs over all
possible dimer coverings having $b$ as the set of free sites. We are
then finished if we can show that the subspace $\mathcal{S}$ given by
Figure \ref{2-star-2}a has trivial intersection with $\mathcal{K}$, or
equivalently, that the sum $\mathcal{S}\oplus \mathcal{K}$ is direct. By
the very definition of the PEPS tensor, $\mathcal{S}$ is given exactly by
$\mathcal{S}=\oplus_b \{\mathcal{H}_b\otimes\sum_{\rm
valid}\hat{Z}|\sigma(D)\>\}$ where $\hat{Z}$ changes the sign of
$|\sigma(D)\>$ when $D$ has one dimer on a link marked by $\bar{Z}$ in
Figure \ref{2-star-2}a and only in this case. By the linear
independence property of Seidel, the sum $\mathcal{S}\oplus \mathcal{H}$
will be direct as long as we can guarantee that for each $b$, in
$\sum_{\rm valid}\hat{Z}|\sigma(D)\>$ there will be terms in the sum with
dimers on the $\bar{Z}$-links and terms without dimers on the
$\bar{Z}$-links. This last statement can be easily checked by ``completing
the path'', as illustrated in Figure \ref{2-star-2}b. The fact that for
each $b$, there exists a Hamiltonian-cycle of alternating colors for the
sites not in $b$ can be easily seen as follows. We can have paths of
alternating colors in the desired vertices in all but the middle bow-tie.
Since $b$ contains an odd number of sites, at one side of the bow tie we
arrive with the same color whereas at the other side we arrive with
different colors. Exactly as in Figure \ref{2-star-2}b we can complete
the cycle within the bow-tie.

\end{document}